\documentstyle[12pt]{article}   

\large
\newcommand{\beq}{\begin{equation}}
\newcommand{\eeq}{\end{equation}}

\makeatletter
\@addtoreset{equation}{section}
\makeatother

\newtheorem{Definition}{Definition}[section]
\newtheorem{Lemma}{Lemma}[section]

\def\be{\begin{equation}}
\def\ee{\end{equation}}
\def\ba{\begin{eqnarray}}
\def\ea{\end{eqnarray}}

\def\ag{{{\cal A}/{\cal G}}}

\def\agb{{\overline {{\cal A}/{\cal G}}}}
\def\ab{{\overline{{\cal A}}}}

\def\Comp{{\mathchoice
{\setbox0=\hbox{$\displaystyle\rm C$}\hbox{\hbox to0pt
{\kern0.4\wd0\vrule height0.9\ht0\hss}\box0}}
{\setbox0=\hbox{$\textstyle\rm C$}\hbox{\hbox to0pt
{\kern0.4\wd0\vrule height0.9\ht0\hss}\box0}}
{\setbox0=\hbox{$\scriptstyle\rm C$}\hbox{\hbox to0pt
{\kern0.4\wd0\vrule height0.9\ht0\hss}\box0}}
{\setbox0=\hbox{$\scriptscriptstyle\rm C$}\hbox{\hbox to0pt
{\kern0.4\wd0\vrule height0.9\ht0\hss}\box0}}}}
\def\Co{{\mathchoice
{\setbox0=\hbox{$\displaystyle\rm C$}\hbox{\hbox to0pt
{\kern0.4\wd0\vrule height0.9\ht0\hss}\box0}}
{\setbox0=\hbox{$\textstyle\rm C$}\hbox{\hbox to0pt
{\kern0.4\wd0\vrule height0.9\ht0\hss}\box0}}
{\setbox0=\hbox{$\scriptstyle\rm C$}\hbox{\hbox to0pt
{\kern0.4\wd0\vrule height0.9\ht0\hss}\box0}}
{\setbox0=\hbox{$\scriptscriptstyle\rm C$}\hbox{\hbox to0pt
{\kern0.4\wd0\vrule height0.9\ht0\hss}\box0}}}}
\def\Rl{{\mathchoice
{\setbox0=\hbox{$\displaystyle\rm R$}\hbox{\hbox to0pt
{\kern0.4\wd0\vrule height0.9\ht0\hss}\box0}}
{\setbox0=\hbox{$\textstyle\rm R$}\hbox{\hbox to0pt
{\kern0.4\wd0\vrule height0.9\ht0\hss}\box0}}
{\setbox0=\hbox{$\scriptstyle\rm R$}\hbox{\hbox to0pt
{\kern0.4\wd0\vrule height0.9\ht0\hss}\box0}}
{\setbox0=\hbox{$\scriptscriptstyle\rm R$}\hbox{\hbox to0pt
{\kern0.4\wd0\vrule height0.9\ht0\hss}\box0}}}}

\def\p{p}

\def\dprime{{\prime\prime}}

\title{Closed formula for the matrix elements of the volume operator in
canonical quantum gravity} 
\author{T. Thiemann\thanks{thiemann@math.harvard.edu} \\
       Physics Department, Harvard University, \\
       Cambridge, MA 02138, USA}
\date{{\small Preprint HUTMP-96/B-353}}

\begin{document}

\maketitle

\begin{abstract}
We derive a closed formula for the matrix elements of the volume operator
for canonical Lorentzian quantum gravity in four spacetime dimensions in 
the continuum in a spin-network basis.

We also display a new technique of regularization which is state dependent
but we are forced to it in order to maintain diffeomorphism covariance
and in that sense it is natural.

We arrive naturally at the expression for the volume operator as defined by 
Ashtekar and Lewandowski up to a state independent factor.
\end{abstract}

\section{Introduction}

The volume functional $V$ for Lorentzian canonical quantum gravity plays a 
quite
important role in the quantization process : if one follows the approach 
advertized in \cite{1} and chooses a representation in which traces of 
the holonomy along closed loops for a certain $SU(2)$ $A_a^i$ connection 
can be 
promoted to basic configuration operators, then it becomes technically 
very hard to define physically interesting operators such as the 
Euclidean and Lorentzian Hamiltonian constraint operators, an operator
corresponding to the length of a curve, an operator corresponding to the ADM
energy, matter Hamiltonian operators and so forth. The common reason is 
that all these operators involve only the co-triad $e_a^i$ of the intrinsic 
metric of the initial data hypersurface $\Sigma$ polynomially which, 
however, cannot be written 
as a polynomial in the momentum conjugate to $A_a^i$ and therefore does not
seem to make any sense as a quantum operator.

Fortunately, there is a trick available : as proved in \cite{2,3}, the 
total volume of $\Sigma$ is the generating functional of the co-triad
meaning that one gets $e_a^i$ from $V$ essentially by functional 
derivation. Therefore, if we write $\delta t\dot{s}^a e_a^i\propto
\{h_s,V\}+o(\delta t^2)$ where $s$ is an edge with parameter length 
$\delta t$, $h_s$ is the holonomy  along $s$ and $\{.,.\}$ denotes Poisson
brackets then it would be sufficient to quantize $V$ in such a way that
(gauge-invariant) functions of a finite number of holonomies form a dense 
domain. Such a satisfactory quantization of $V$ which has precisely this 
property was indeed given already in the literature \cite{4,5}. 

Therefore, the strategy outlined above and in \cite{2,3} can be 
implemented and, in particular, was successfully employed to 
construct various operators listed above \cite{2,3,6,7,8}. However, the 
explicit computation of the spectrum of most of those operators was not 
performed so far. The reason
is that the spectra of all these operators are clearly largely determined 
by the spectrum of the volume operator whose precise spectrum is fairly
unknown except for the most simple situations \cite{9,10}.

The present paper is devoted to the derivation of the complete set 
of matrix elements of this operator in closed form from which the spectrum 
can be obtained straightforwardly.

In section 3, after introducing the notation in section 2, we develop a 
novel technique to regularize the volume operator which is state dependent
very much in the same way as the regularization displayed in the second
reference of \cite{5}. This part of the paper is a continuation of the 
work on the point splitting regularizations of the volume and area operators
as displayed in \cite{1,5}.
The derivation is rather short and gives a diffeomorphism covariant end 
result.
We are unambiguously led to the Ashtekar-Lewandowski volume operator 
defined in \cite{18} up to a factor of $\sqrt{27/8}$. 

In section 4 we repeat the argument given in \cite{5,9} which shows that 
the spectrum is discrete and that its computation reduces to linear 
algebra, in the sense that we just need to determine 
separately the eigenvalues of an infinite number of finite-dimensional, 
positive semi-definite and symmetric real-valued matrices.

Finally, in section 5 we derive the explicit formula for an arbitrary 
matrix element in a spin-network basis and compute the eigenvalues of the 
associated matrices in a number of simple cases. It should be noted that
the formulae obtained, albeit quite complicated, are the starting point 
for a numerical evaluation of the spectrum and are in particular useful to
determine the large spin behaviour of the volume operator.

\section{Notation}

Let us first introduce the notation. We consider the Hamiltonian 
formulation of the Palatini action of four-dimensional Lorentzian 
vacuum gravity.\\ 
Denote by $q_{ab}$ the intrinsic 
metric on the initial data hypersurface $\Sigma$ and let $e_a^i$ be
its co-triad where $i,j,k,..$ denote $su(2)$ indices. The densitized 
triad $E^a_i:=\det((e_b^j))e^a_i$ multiplied by $1/\kappa,\;\kappa$ being
Newton's constant, is the momentum 
conjugate to $K_a^i:=\mbox{sgn}(\det((e_c^j)))K_{ab}e^b_i$ where 
$K_{ab}$ is the extrinsic curvature of $\Sigma$. Upon performing a canonical
point transformation \cite{11,12} on the phase space coordinatized by 
$(K_a^i,E^a_i/\kappa)$ we arrive at the chart 
$(A_a^i/\kappa:=(\Gamma_a^i+K_a^i)/\kappa,E^a_i)$ where $\Gamma_a^i$ is the 
spin-connection of $e_a^i$. It is easy to see that the configuration 
variable $A_a^i$ is a $su(2)$ connection.

The virtue of casting general relativity into a connection dynamics 
formulation is as follows.
\begin{itemize}
\item[1)] First of all, the phase space of general relativity is now embedded
into that of an $su(2)$ gauge theory, thereby opening access to a wealth of
techniques that have been proving useful in quantizing those theories. 
Probably the most important technique is the use of traces of the 
holonomy along piecewise analytic loops in $\Sigma$, so-called Wilson-loop 
functionals, to 
coordinatize the space of smooth connections modulo gauge transformations
$\ag$ \cite{13,14}.
\item[2)] Using those Wilson loop functionals one can construct 
\cite{15,16,17} a quantum configuration space of generalized connection 
modulo gauge transformations $\agb$ and find natural, $\sigma$-additive,
diffeomorphism invariant, regular Borel probability measures $\mu_0$
thereon which equips us with a Hilbert space structure 
${\cal H}:=L_(\agb,d\mu_0)$. Of course, for this to make sense we have to 
make the following assumtion : Wilson loop functionals continue to make
sense as operators in the quantum theory. We will adopt this viewpoint 
in the sequel in the hope to capture one of the physically interesting 
phases of quantum general relativity. The interested reader is referred to
\cite{1} and references therein for further details.
\end{itemize}
The following paragraph is supposed to equip the reader with a working 
knowledge of the tools associated with $\agb$.\\
We begin by defining cylindrical functions. Since the gauge invariant 
information contained in a connection is captured by finite linear 
sums of products of Wilson loop functionals we may label gauge 
invariant functions by piecewise analytic graphs $\gamma$ which are just the 
union of the loops involved. Such a graph consists of maximally 
analytic pieces, called the edges $e$ of $\gamma$ and the edges
meet in the vertices $v$ of $\gamma$. A function $f$ on $\agb$ is said 
to be cylindrical with respect to a graph $\gamma$ iff it can be written
as $f=f_\gamma\circ\p_\gamma$ where $p_\gamma(A)=(h_{e_1}(A),..,h_{e_n}(A))$
and where $e_1,..,e_n$ are the edges of $\gamma$. Here $h_e(A)$ is the 
holonomy along $e$ evaluated at $A\in\agb$ and $f_\gamma$ is a complex
valued function on $SU(2)^n$. Since a function cylindrical with respect 
to a graph $\gamma$ is automatically cylindrical with respect to any 
graph bigger than $\gamma$, a cylindrical function is actually given 
by a whole equivalence class of functions $f_\gamma$. In the sequel we 
will not distinguish between this equivalence class and one of its 
representants.\\
We say that a function cylindrical with respect to a graph $\gamma$ is
of class $C^n$ if it is of class $C^n$ with respect to the 
standard differentiable structure of $SU(2)^n$ and denote this class
of functions by $\mbox{Cyl}_\gamma^n(\agb),\;\mbox{Cyl}^n(\agb):=
\cup_{\gamma\in\Gamma}\mbox{Cyl}_\gamma^n(\agb)$ where $\Gamma$ is the 
set of piecewise analytic graphs in $\Sigma$. Thus we have a differential
calculus \cite{18} on $\agb$.\\
Finally, the measure $\mu_0$ is the faithful projective limit \cite{19,20}
of the self-consistent projective family $(\mu_{0,\gamma})_{\gamma\in\Gamma}$
defined by
$$
\int_\agb d\mu_0(A) f(A):=\int_\agb d\mu_{0,\gamma}(A) 
f_\gamma(p_\gamma(A)):=\int_{SU(2)^n} d\mu_H(g_1)..d\mu_H(g_n) 
f_\gamma(g_1,..,g_n)
$$
where $g_I:=h_{e_I}(A)$. This equips us with an integral calculus on $\agb$.

\section{Regularization of the volume operator}

In this section we present a new regularization of the volume operator
which uses the methods derived by Abhay Ashtekar and Jurek Lewandowski
in \cite{18,5,9}. 
Most of the issues involved in the regularization as presented here are 
borrowed from their work.\\
The advantage of our procedure is that it is rather short and compact.\\
\\
Let $R\subset\Sigma$ be an open, connected region of $\Sigma$. 
We have the identity
\be \label{1}
\frac{1}{3!}\epsilon_{abc}\epsilon^{ijk}E^a_i E^b_j E^c_k 
=\det((E^a_i))=\det((q_{ab}))=[\det((e_a^i))]^2=:\det(q)\ge 0
\ee
and can thus write the volume of the region $R$ as measured by the metric
$q_{ab}$ as follows
\be \label{2}
V(R):=\int_R d^3x \sqrt{\det(q)}=
\sqrt{\frac{1}{3!}\epsilon_{ijk}\epsilon^{abc}E^a_i E^b_j E^c_k}\;.
\ee
The next step is to smear the fields $E^a_i$. Let $\chi_\Delta(p,x)$ be 
the characteristic function in the coordinate $x$ of a cube with 
center $p$ spanned by the three vectors 
$\vec{\Delta}_i=\Delta_i\vec{n}_i(\Delta)$ 
where $\vec{n}_i$ is a normal vector in the frame under consideration 
and which has coordinate volume $\mbox{vol}=\Delta_1\Delta_2\Delta_3
\det(\vec{n}_1,\vec{n}_2,\vec{n}_3)$ (we assume the three normal vectors to
be right oriented). In other words, $\chi_\Delta(p,x)=
\prod_{i=1}^3\theta(\Delta_i-|<n_i,x-p>|)$ where $<.,.>$ is the standard 
Euclidean inner product and $\theta(y)=1$ for $y>0$ and zero otherwise.\\
We consider the smeared quantity 
\ba \label{3}
&& E(p,\Delta,\Delta',\Delta^\dprime):=
\frac{1}{\mbox{vol}(\Delta)\mbox{vol}(\Delta')\mbox{vol}(\Delta^\dprime)}
\int_\Sigma d^3x\int_\Sigma d^3y \int_\Sigma d^3z 
\chi_\Delta(p,x)\chi_{\Delta'}(p,\frac{x+y}{2})\times\nonumber\\
&&\times \chi_{\Delta^\dprime}(p,\frac{x+y+z}{3}) 
\frac{1}{3!}\epsilon_{abc}\epsilon^{ijk}E^a_i(x) E^b_j(y) E^c_k(z)\;.
\ea 
Notice that if we take the limits $\Delta_i,\Delta_i',\Delta_i^\dprime\to 0$ 
in any
combination and in any rate with respect to each other then we get back to 
(\cite{1}) evaluated at the point 
$p$. This holds for any choice of linearly independent normal
vectors $\vec{n}_i,\vec{n}_i',\vec{n}_i^\dprime$.
The strange argument $(x+y+z)/3$ will turn out to be very crucial in 
obtaining
a manifestly diffeomorphism covariant result. We will see this in a moment.\\
Then it is easy to see that the classical identity
\be \label{4}
V(R)=\lim_{\Delta\to 0} \lim_{\Delta'\to 0}\lim_{\Delta^\dprime\to 0}
\int_R d^3p \sqrt{|E(p,\Delta,\Delta',\Delta^\dprime)|}
\ee
holds. We could introduce absolute values because we have the classical
identity $\det(q)=|\det(q)|$ and the modulus is necessary for the 
regulated quantity (\ref{3}) is not non-negative anymore.\\
The virtue of introducing the quantities (\ref{3}) is that they can
be promoted to quantum operators which have the dense domain 
$\mbox{Cyl}^1(\agb)$. To see this, notice that due to the canonical 
brackets
$\{A_a^i(x),E^b_j(y)\}=\kappa\delta_a^b\delta_j^i\delta^{(3)}(x,y)$
we are naturally led to represent the operator corresponding 
to $E^a_i$ by $\hat{E}^a_i(x)=-i\ell_p^2\delta/\delta A_a^i(x)$ where
$\ell_p=\sqrt{\hbar\kappa}$ is the Planck length. \\
Now let be given a graph
$\gamma$. In order to simplify the notation, we subdivide each edge 
$e$ with endpoints $v,v'$ which are vertices of $\gamma$ into two segments
$s,s'$ where $e=s\circ (s')^{-1}$ and $s$ has an orientation 
such that it is {\em outgoing} at $v$ while $s'$ has an orientation 
such that it is {\em outgoing} at $v'$. This introduces new vertices 
$s\cap s'$ which we will call pseudo-vertices because they are not 
points of non-analyticity of the graph. Let $E(\gamma)$ be the set of
these segments of $\gamma$ but $V(\gamma)$ the set of true (as opposed to 
pseudo) vertices of $\gamma$. Let us now evaluate the action of
$\hat{E}^a_i(p,\Delta):=1/\mbox{vol}(\Delta)\int_\Sigma d^3x 
\chi_\Delta(p,x)\hat{E}^a_i(x)$ on a function $f=p_\gamma^\ast f_\gamma$ 
cylindrical with respect to $\gamma$. We find ($e:\;[0,1]\to\Sigma;\;
t\to e(t)$ being a parametrization of the edge $e$) 
\ba \label{5}
&& \hat{E}^a_i(p,\Delta)f=-\frac{i\ell_p^2}{\mbox{vol}(\Delta)}
\sum_{e\in E(\gamma)} \int_0^1 dt \chi_\Delta(p,e(t))
\dot{e}^a(t) \times \nonumber\\
& \times & \frac{1}{2} \mbox{tr}(h_e(0,t)\tau_i 
h_e(t,1)\frac{\partial}{\partial h_e(0,1)})f_\gamma \;.
\ea
Here we have used 1) the fact that a cylindrical function is 
already determined by
its values on $\ag$ rather than $\agb$ so that it makes sense to take the
functional derivative, 2) the definition of the holonomy as the path ordered 
exponential of $\int_e A$ with the smallest parameter value to the left,
3) $A=dx^a A_a^i \tau_i/2$ where $SU(2)\ni\tau_j=-i\sigma_j,\;\sigma_i$ 
being the usual Pauli matrices, so that $[\tau_i/2,\tau_j/2]=\epsilon_{ijk}
\tau_k/2$ and 4) we have defined $\mbox{tr}(h\partial/\partial g)=
h_{AB}\partial/\partial g_{AB},\; A,B,C,..$ being $SU(2)$ indices.\\
We now wish to evaluate the whole operator 
$\hat{E}(p,\Delta,\Delta',\Delta^\dprime)$ on $f$. 
It is clear that we obtain three types of terms, the first type 
comes from all three functional derivatives acting on $f$ only, the second
type comes from two functional derivatives acting on $f$ and the remaining 
one acting on the trace appearing in (\ref{5}) and finally the third type
comes from only one derivative acting on $f_\gamma$ and the remaining
two acting on the trace. Explicitly we find (we mean by $\theta(t,t')$
the theta function which is unity if $0<t<t'<1$, $1/2$ if 
$t=0<t'<1,0<t<t'=0$ and $1/4$ if $t\le t'\in\{0,1\}$ and zero otherwise.
Likewise $\theta(t,t',t^\dprime)$ is $1$ if $t<t'<t^\dprime$, $1/2$ if
$t=0<t'<t^\dprime,0<t<t'<t^\dprime=1$, $1/4$ if
$0=t=t'<t^\dprime,0<t<t'=t^\dprime$ and $1/8$ if $t\le t'\le 
t^\dprime\in\{0,1\}$. Finally $m(t,t',t^\dprime)=1/2^k$ where $k=0,1,2,3$
is the possible number of arguments that equal $0,1$) 
\ba \label{6} 
&& \hat{E}(p,\Delta,\Delta',\Delta^\dprime)f
=\frac{i\ell_p^6}{8\cdot 3!
\mbox{vol}(\Delta)\mbox{vol}(\Delta')\mbox{vol}(\Delta^\dprime)}
\epsilon_{abc}\epsilon^{ijk}\int_{[0,1]^3}dt dt' dt^\dprime\times\nonumber\\
& \times& \left\{ \sum_{e,e',e^\dprime\in E(\gamma)}
\dot{e}(t)^a\dot{e}'(t')^b\dot{e}^\dprime(t^\dprime)^c
\chi_\Delta(p,e(t))\chi_{\Delta'}(p,\frac{e(t)+e'(t')}{2})
\right.\times\nonumber\\
&\times&
\chi_{\Delta^\dprime}(p,\frac{e(t)+e'(t')+e^\dprime(t^\dprime)}{3})
m(t,t',t^\dprime)\mbox{tr}(h_{e^\dprime}(0,t^\dprime)\tau_k 
h_{e^\dprime}(t^\dprime,1)
\frac{\partial}{\partial h_{e^\dprime}(0,1)}) \times\nonumber\\
&\times&
\mbox{tr}(h_{e'}(0,t')\tau_j h_{e'}(t',1)
\frac{\partial}{\partial h_{e'}(0,1)})
\mbox{tr}(h_e(0,t)\tau_i h_e(t,1)\frac{\partial}{\partial h_e(0,1)}) 
\nonumber\\
&+&
\sum_{e',e^\dprime\in E(\gamma)}
\dot{e}^\dprime(t)^a\dot{e}'(t')^b\dot{e}^\dprime(t^\dprime)^c
\chi_\Delta(p,e^\dprime(t))\chi_{\Delta'}(p,\frac{e^\dprime(t)+e'(t')}{2})
\times\nonumber\\
&\times&
\chi_{\Delta^\dprime}(p,\frac{e^\dprime(t)+e'(t')+e^\dprime(t^\dprime)}{3})
[\theta(t,t^\dprime)\mbox{tr}(h_{e^\dprime}(0,t)\tau_i 
h_{e^\dprime}(t,t^\dprime) \tau_k h_{e^\dprime}(t^\dprime,1)
\frac{\partial}{\partial h_{e^\dprime}(0,1)}) 
\nonumber\\
&& +\theta(t^\dprime,t)\mbox{tr}(h_{e^\dprime}(0,t^\dprime)\tau_k 
h_{e^\dprime}(t^\dprime,t) 
\tau_i h_{e^\dprime}(t,1)\frac{\partial}{\partial h_{e^\dprime}(0,1)}) ]
\mbox{tr}(h_{e'}(0,t')\tau_j h_{e'}(t',1)
\frac{\partial}{\partial h_{e'}(0,1)})
\nonumber\\
&+&
\sum_{e',e^\dprime\in E(\gamma)}
\dot{e}'(t)^a\dot{e}'(t')^b\dot{e}^\dprime(t^\dprime)^c
\chi_\Delta(p,e'(t))\chi_{\Delta'}(p,\frac{e'(t)+e'(t')}{2})
\times\nonumber\\
&\times&
\chi_{\Delta^\dprime}(p,\frac{e'(t)+e'(t')+e^\dprime(t^\dprime)}{3})
\mbox{tr}(h_{e^\dprime}(0,t^\dprime)\tau_k h_{e^\dprime}(t^\dprime,1)
\frac{\partial}{\partial h_{e^\dprime}(0,1)}) 
\times\nonumber\\
&\times&
[\theta(t,t')\mbox{tr}(h_{e'}(0,t)\tau_i h_{e'}(t,t') 
\tau_j h_{e'}(t',1)\frac{\partial}{\partial h_{e'}(0,1)}) 
\nonumber\\
&& +
\theta(t',t)\mbox{tr}(h_{e'}(0,t')\tau_j 
h_{e'}(t',t)\tau_i h_{e'}(t,1)\frac{\partial}{\partial h_{e'}(0,1)}) ]
\nonumber\\
&+&
\sum_{e,e^\dprime\in E(\gamma)}
\dot{e}(t)^a\dot{e}^\dprime(t')^b\dot{e}^\dprime(t^\dprime)^c
\chi_\Delta(p,e(t))\chi_{\Delta'}(p,\frac{e(t)+e^\dprime(t')}{2})
\times\nonumber\\
&\times&
\chi_{\Delta^\dprime}(p,\frac{e(t)+e^\dprime(t')+e^\dprime(t^\dprime)}{3})
[\theta(t',t^\dprime)\mbox{tr}(h_{e^\dprime}(0,t')\tau_j 
h_{e^\dprime}(t',t^\dprime) \tau_k h_{e^\dprime}(t^\dprime,1)
\frac{\partial}{\partial h_{e^\dprime}(0,1)}) 
\nonumber\\ 
&& +
\theta(t^\dprime,t')\mbox{tr}(h_{e^\dprime}(0,t^\dprime)\tau_k 
h_{e^\dprime}(t^\dprime,t') 
\tau_j h_{e^\dprime}(t',1)\frac{\partial}{\partial h_{e^\dprime}(0,1)}) ]
\mbox{tr}(h_e(0,t)\tau_i h_e(t,1)
\frac{\partial}{\partial h_e(0,1)})
\nonumber\\
&+&
\sum_{e^\dprime\in E(\gamma)}
\dot{e}^\dprime(t)^a\dot{e}^\dprime(t')^b\dot{e}^\dprime(t^\dprime)^c
\chi_\Delta(p,e^\dprime(t))
\chi_{\Delta'}(p,\frac{e^\dprime(t)+e^\dprime(t')}{2})
\times\nonumber\\
&\times&
\chi_{\Delta^\dprime}
(p,\frac{e^\dprime(t)+e\dprime(t')+e^\dprime(t^\dprime)}{3})
\times\nonumber\\
&\times&
[\theta(t,t',t^\dprime)\mbox{tr}(h_{e^\dprime}(0,t)\tau_i 
h_{e^\dprime}(t,t') 
\tau_j h_{e^\dprime}(t',t^\dprime)\tau_k 
h_{e^\dprime}(t^\dprime,1)\frac{\partial}{\partial h_{e^\dprime}(0,1)}) 
\nonumber\\ && +
\theta(t,t^\dprime,t')\mbox{tr}(h_{e^\dprime}(0,t)\tau_i 
h_{e^\dprime}(t,t^\dprime) 
\tau_k h_{e^\dprime}(t^\dprime,t')\tau_j 
h_{e^\dprime}(t',1)\frac{\partial}{\partial h_{e^\dprime}(0,1)}) 
\nonumber\\
&& +
\theta(t',t^\dprime,t)\mbox{tr}(h_{e^\dprime}(0,t')\tau_j 
h_{e^\dprime}(t',t^\dprime) 
\tau_k h_{e^\dprime}(t^\dprime,t)\tau_i 
h_{e^\dprime}(t,1)\frac{\partial}{\partial h_{e^\dprime}(0,1)}) 
\nonumber\\ && +
\theta(t',t,t^\dprime)\mbox{tr}(h_{e^\dprime}(0,t')\tau_j 
h_{e^\dprime}(t',t) 
\tau_i h_{e^\dprime}(t,t^\dprime)\tau_k 
h_{e^\dprime}(t^\dprime,1)\frac{\partial}{\partial h_{e^\dprime}(0,1)}) 
\nonumber\\ && +
 \theta(t^\dprime,t,t')\mbox{tr}(h_{e^\dprime}(0,t^\dprime)\tau_k 
h_{e^\dprime}(t^\dprime,t) 
\tau_i h_{e^\dprime}(t,t')\tau_j 
h_{e^\dprime}(t',1)\frac{\partial}{\partial h_{e^\dprime}(0,1)}) 
\nonumber\\ && + \left.
\theta(t^\dprime,t',t)\mbox{tr}(h_{e^\dprime}(0,t^\dprime)\tau_k 
h_{e^\dprime}(t^\dprime,t') 
\tau_j h_{e^\dprime}(t',t)\tau_i 
h_{e^\dprime}(t,1)\frac{\partial}{\partial h_{e^\dprime}(0,1)})]
\right\} f_\gamma \nonumber\\
&=:& 
[\hat{O}_{1,2,3}+\hat{O}_{2,31}+\hat{O}_{12,3}+\hat{O}_{1,23}+\hat{O}_{123}]
f_\gamma \;. 
\ea
The fact that the integrand of the terms involved in 
$\hat{O}_{12,3},\hat{O}_{1,23},\hat{O}_{2,31},\hat{O}_{123}$
vanishes if either of the cases 
$0<t=t'<1,0<t'=t^\dprime<1,0<t=t'=t^\dprime<1$ occurs is due to the fact 
that in this case in
$\hat{O}_{12,3},\hat{O}_{1,23},\hat{O}_{2,31},\hat{O}_{123}$ we get
a trace which contains $\tau_{(i}\tau_{j)},\tau_{(j}\tau_{k)},
\tau_{(k}\tau_{i)}$ contracted with
$\epsilon^{ijk}$ which vanishes (to see this recall that the functional
derivative is 
\ba \label{6a}
&& \delta h_e(A)/\delta A_a^i(x)=
\frac{1}{2}\int_0^1 dt 
[\frac{1}{2}\delta^{(3)}(e(t+),x)\dot{e}(t+)^a h_e(0,t)\tau_i h_e(t,1)
\nonumber\\
&& +\frac{1}{2}\delta^{(3)}(e(t-),x)\dot{e}(t-)^a h_e(0,t)\tau_i h_e(t,1)]
\nonumber
\ea
(one sided derivatives and $\delta$ distributions). This expression is 
also correct if $x$ is an endpoint of $e$ (in which case there is only
one term which survives in (\ref{6a})).
which 
results in the case that we consider $h_{e_1}\tau_j h_{e_2}$
instead of $h_e, e=e_1\circ e_2,x=e_1\cap e_2$ a point of analyticity, in a 
term involving $\tau_{(i}\tau_{j)}$).\\
Given a triple $e,e',e^\dprime$ of (not necessarily distinct) 
edges of $\gamma$, consider the functions 
\be \label{7}
x_{e e' e^\dprime}(t,t',t^\dprime):=\frac{e(t)+e'(t')
+e^\dprime(t^\dprime)}{3}\;. \ee
This function has the interesting property that the Jacobian is given by
\be \label{8}
3^3\det(\frac{\partial(x^1_{e e' e^\dprime},x^2_{e e' e^\dprime},
x^3_{e e' e^\dprime})(t,t',t^\dprime)}
{\partial(t,t',t^\dprime)})=\epsilon_{abc}
\dot{e}(t)^a\dot{e}'(t')^b\dot{e}^\dprime(t^\dprime)^a
\ee
which is precisely the form of the factor which enters all the integrals 
in (\ref{6}). This is why we have introduced the strange argument 
$(x+y+z)/3$.\\
We now consider the limit $\Delta_i,\Delta'_i,\Delta^\dprime_i\to 0$. 
\begin{Lemma} \label{la1}
For each triple of edges $e,e',e^\dprime$ there exists a choice of vectors 
$\vec{n}_i,\vec{n}_i',\vec{n}_i^\dprime$ and a way to 
guide the limit $\Delta_i,\Delta'_i,\Delta^\dprime_i\to 0$ such that
\be \label{6b}
\int_{[0,1]^3} \det(\frac{\partial(x^a_{e e' 
e^\dprime})}{\partial(t,t',t^\dprime)}) 
\chi_\Delta(p,e)\chi_{\Delta'}(p,(e+e')/2)
\chi_{\Delta^\dprime}(p,(e+e'+e^\dprime)/3)\hat{O}_{e e' e^\dprime}
\ee
vanishes\\ 
a) if $e,e',e^\dprime$ do not all intersect $p$ or\\ 
b)
$\det(\frac{\partial(x^a_{e e' e^\dprime})}{\partial(t,t',t^\dprime)})_p=0$
(which is a diffeomorphism invariant statement). \\
Otherwise it tends to\\ $1/8\mbox{sgn}(\det(\frac{\partial(x^a_{e e' 
e^\dprime})}{\partial((t,t',t^\dprime)}))_p 
\hat{O}_{e,e',e^\dprime}(p)\prod_{i=1}^3\Delta_i^\dprime$. Here 
we have denoted by $\hat{O}_{e e' e^\dprime}(t,t',t^\dprime)$ the trace(s) 
involved in the various terms of (\ref{6}).
\end{Lemma}
Proof\footnote{The idea that one should adapt the regularization
to each triple of edges was first communicated to the author by
Jurek Lewandowski. Its justification relies on the fact that the 
classical expression does not depend on the way we regularize. What is new 
here is to introduce even more parameters than in \cite{9} to be adapted 
to a triple which simplifies the proofs. 
What is also new is that we introduced the $(x+y+z)/3$ argument which 
makes the proof especially clear.}  :\\
If at least one of $e,e',e^\dprime$ does not intersect $p$ then, if we choose
$\Delta_i$ etc. smaller than some finite number $\Delta_0$, (\ref{6b})
vanishes identically since the support of the characteristic functions is 
in a neighbourhood around $p$ which shrinks to zero with the $\Delta_i$
etc.\\
So let us assume that all of $e,e',e^\dprime$ intersect $p$ at parameter
value $t_0,t_0',t_0^\dprime$ (this value is unique because the edges are not 
self-intersecting). Then we can write $e(t)=p+c(t-t_0)$ where 
$c$ is analytic and vanishes at $\tau=t-t_0=0$. We have the 
case subdivision :\\ 
Case I : $\det(\frac{\partial(x^a_{e e' 
e^\dprime})}{\partial(t,t',t^\dprime)}))_{t_0}=0$.\\
Case Ia) : All of $\dot{c}(0),\dot{c}'(0),\dot{c}^\dprime(0)$ are 
co-linear.\\ 
Case Ib) : Two of $\dot{c}(0),\dot{c}'(0),\dot{c}^\dprime(0)$ are 
co-linear and the third is linearly independent of them.\\ 
Case Ic) : No two of $\dot{c}(0),\dot{c}'(0),\dot{c}^\dprime(0)$ are 
co-linear.\\
Case II : $\det(\frac{\partial(x^a_{e e' 
e^\dprime})}{\partial(t,t',t^\dprime)}))_{t_0}\not=0$.\\
Notice that all vectors $\dot{c}(0),\dot{c}'(0),\dot{c}^\dprime(0)$ 
are non-vanishing by the definition of a curve.

We consider first case I). We exclude the trivial case that all three 
curves lie in a coordinate plane or line such that the determinant 
already vanishes for all finite values of the $\Delta$'s. Therefore there
exist linearly independent unit vectors $u,v,w$ (not necessarily 
orthogonal) in terms of which we may express $c,c',c^\dprime$.\\ 
In case Ia) we have an expansion of the form 
\ba 
c(t)&=&au(t+o(t^2))+bv(t^m+o(t^{m+1}))+cw(t^n+o(t^{n+1}))\nonumber\\
c'(t)&=&au(t+o(t^2))+b'v(t^{m'}+o(t^{m'+1}))+c'w(t^{n'}+o(t^{n'+1}))\nonumber\\
c^\dprime(t)&=&a^\dprime 
u(t+o(t^2))+b^\dprime 
v(t^{m^\dprime}+o(t^{m^\dprime+1}))+c^\dprime 
w(t^{n^\dprime}+o(t^{n^\dprime+1}))\nonumber\\
\ea
where $a,b,c,a',b',c',a^\dprime,b^\dprime,c^\dprime$ are real numbers with
$a a' a^\dprime\not=0$ and at least one of the $b$'s and $c$'s being 
different from zero (also not for instance $b=c=b'=c'=0$). Furthermore
$m,m',m^\dprime,n,n',n^\dprime\ge 2$.
The characteristic functions have support in coordinate cubes spanned 
by the vectors $\vec{n}_i,\vec{n}_i',\vec{n}_i^\dprime$. Now, since $u,v,w$ 
are linearly independent we may simply {\em choose}, for instance,
$\vec{n}_i:=u,\vec{n}_i':=v,\vec{n}_i^\dprime:=w$.
It follows then and from the fact that $0\le\chi\le 1$ that
\ba \label{9}
&&\chi_\Delta(p,e)\chi_{\Delta'}(p,(e+e')/2)
\chi_{\Delta^\dprime}(p,(e+e'+e^\dprime)/3)\nonumber\\
&=&\tilde{\chi}_{\Delta}(0,c)
\tilde{\chi}_{\Delta'}(0,(c+c')/2)
\tilde{\chi}_{\Delta^\dprime}(0,(c+c'+c^\dprime)/3)\nonumber\\
&\le& 
\theta_{\Delta_1}(<c,u>)
\theta_{\Delta_1'}(<(c+c')/2,v>)
\theta_{\Delta_1^\dprime}(<(c+c'+c^\dprime)/3,w>)\;.
\ea
From the explicit expansions of $c,c',c^\dprime$ we conclude that (\ref{9})
has the bound 
\be \label{10}
\theta_{\delta_1\Delta_1}(t)
\theta_{\delta_1'\Delta_1'}(t')
\theta_{\delta_1^\dprime\Delta_1^\dprime}(t^\dprime)
\ee
for some sufficently large numbers $\delta_1,\delta_1',\delta_1^\dprime$.
On the other hand we also see from the explicit expansion of 
$|\det(\frac{\partial(x^a_{e e' 
e^\dprime})}{\partial(t,t',t^\dprime)}))|$ around $t_0$ that it is bounded by
$M(|t|^k+|t'|^k+|t^\dprime|^k)$ where $M$ is a positive number and where
$k=\min(m+n',m+n^\dprime,m'+n,m'+n^\dprime,m^\dprime+n,m^\dprime+n')-2
\ge 2$.\\
The prescription of how to guide the limit in case Ia) is then to
synchronize $\Delta_1=\Delta_1'=\Delta_1^\dprime=\Delta$ and to take the 
limit $\Delta\to 0$ first. The integral is at least of order $\Delta^5$
while we divide only by an order of $\Delta^3$ so that the result vanishes.\\
In case Ib) we have an expansion of the form (let w.l.g. $c,c'$ have 
co-linear tangents)
\ba 
c(t)&=&au(t+o(t^2))+bv(t^m+o(t^{m+1}))+cw(t^n+o(t^{n+1}))\nonumber\\
c'(t)&=&au(t+o(t^2))+b'v(t^{m'}+o(t^{m'+1}))+c'w(t^{n'}+o(t^{n'+1}))\nonumber\\
c^\dprime(t)&=&a^\dprime 
v(t+o(t^2))+b^\dprime 
u(t^{m^\dprime}+o(t^{m^\dprime+1}))+c^\dprime 
w(t^{n^\dprime}+o(t^{n^\dprime+1})).\nonumber\\
\ea
We now argue as above and find that the product of the characteristic
functions can be estimated by
$$
\theta_{\delta_1\Delta_1}(t)
\theta_{\delta_1'\Delta_1'}(t')
\theta_{\delta_2^\dprime\Delta_2^\dprime}(t^\dprime)
$$
while the determinant can be estimated as above just that $k$ is now given
by $k=\min(m,m',m^\dprime,n,n',n^\dprime)-1\ge 1$.\\
The prescription is now $\Delta_1=\Delta_1'=\Delta_2^\dprime=:\Delta
\to 0$ first and we conclude that the integral is at least of order 
$\Delta^4$ while we divide again only by $\Delta^3$ such that the limit
vanishes.\\
In case Ic) finally we have an expansion of the form  
\ba 
c(t)&=&au(t+o(t^2))+bv(t^m+o(t^{m+1}))+cw(t^n+o(t^{n+1}))\nonumber\\
c'(t)&=&av(t+o(t^2))+b'v(t^{m'}+o(t^{m'+1}))+c'w(t^{n'}+o(t^{n'+1}))\nonumber\\
c^\dprime(t)&=&a^\dprime 
u(t+o(t^2))+b^\dprime 
v(t+o(t^2))+c^\dprime 
w(t^{n^\dprime}+o(t^{n^\dprime+1})).\nonumber\\
\ea
This time we estimate the product of the characteristic functions for 
instance by
$$
\theta_{\delta_1\Delta_1}(t)
\theta_{\delta_2'\Delta_2'}(t')
\theta_{\delta_2^\dprime\Delta_2^\dprime}(t^\dprime)
$$
while the determinant can be estimated as above and  $k$ is given
by $k=\min(m,m',n,n',n^\dprime)-1\ge 1$ so that we have actually
the same situation as in case Ib) upon synchronizing this time
$\Delta_1=\Delta_2'=\Delta_2^\dprime=:\Delta\to 0$.

As for case II) we observe that the nonvanishing of the functional 
determinant at $p$ implies that the map $x_{e e' e^\dprime}$ is actually
invertible in a neighbourhood of $p$ by the inverse function theorem. 
In other words, there is only one point $(t_0,t_0',t_0^\dprime)$ such that
$x_{e e' e^\dprime}(t_0,t_0',t_0^\dprime)=p$. Moreover, since the determinant
is non-vanishing at $p$, all three edges must be distinct form each other.
It follows now from our choice of edges that $p$ must be a vertex 
$v=e\cap e'\cap e^\dprime$ of 
$\gamma$ in order that he result is non-vanishing and thus from the choice of 
parametrization $t_0=t_0'=t_0^\dprime=0$.\\ 
Therefore, if we take 
the limit $\Delta_i^\dprime\to 0$ first in any order then the condition 
$\chi_{\Delta^\dprime}(p,x_{e e' e^\dprime})=1$ will actually imply
$\chi_\Delta(p,e)=\chi_{\Delta'}(p,(e+e')/2)=1$ for small enough
$\Delta_i^\dprime$ so that we can take these characteristic functions out of 
the integral and replace them by $1$ if $p$ is a common vertex of all 
three edges. Also we can replace
the operator $\hat{O}_{e e' e^\dprime}(t,t',t^\dprime)$ by
$\hat{O}_{e e' e^\dprime}(v)$. This holds only if the triple intersects in 
$p$.\\ 
If not all of $e,e',e^\dprime$
intersect in $p$ then the limit will vanish anyway if we take a suitable 
limit of the $\Delta_i$ as we have shown before. We can account for that 
case by replacing $\chi_\Delta(p,e),\chi_{\Delta'}(p,(e+e')/2)$ by 
$\chi_\Delta(p,v)\chi_{\Delta'}(p,v)$. Here $v$ is the common vertex 
at which the distinct $e,e',e^\dprime$ must be incident 
otherwise they could not even pass through a small enough neighbourhood of 
$p$. We can also assume that all three edges have linearly independent
tangents at $v$ and expand still around $t=0$. 
The remaining integral divided
by $\Delta_1^\dprime\Delta_2^\dprime\Delta_3^\dprime$ then tends to
\be \label{11}
\int_{[0,1]^3}d^3t \det(\frac{\partial x_{e e' e^\dprime}}{\partial t})
\delta^{(3)}(p,x_{e e' e_\dprime})
=s(e,e',e^\dprime)\int_{C_{e e' e^\dprime}}d^3x \delta^{(3)}(p,x)
=\frac{1}{8} s(e,e',e^\dprime)
\ee
where
\be \label{12}
s(e,e',e^\dprime)_v:=\mbox{sgn}(\det(\dot{e}(0),\dot{e}'(0),
\dot{e}^\dprime(0))).
\ee
The factor $1/8$ is due to the fact that in the limit
$\Delta^\dprime\to 0$ we obtain an integral over
$C(e,e',e^\dprime)$, the cone based at $p$ and spanned by 
$\dot{e}(0),\dot{e}'(0),\dot{e}^\dprime(0)$ where the orientation is taken
to be positive. This integral just equals $\int_{\Rl_+^3}d^3t 
\delta(0,t)\delta(0,t')\delta(0,t^\dprime)=1/8$ as one can easily check.
This furnishes the proof.\\
$\Box$\\
We conclude that (\ref{6}) reduces to (in particular, the operators
$\hat{O}_{12,3}\hat{O}_{1,23}\hat{O}_{2,31}\hat{O}_{123}$ drop out)
$$
\lim_{\Delta^\dprime\to 0} \hat{E}(p,\Delta,\Delta',\Delta^\dprime)f
=\sum_{e,e',e^\dprime}\frac{i\ell_p^6s(e,e',e^\dprime)_v}{8\cdot 
3!\mbox{vol}(\Delta) \mbox{vol}(\Delta')}
\chi_\Delta(p,v)\chi_{\Delta'}(p,v)
\frac{3^3}{8}\hat{O}_{e,e',e^\dprime}(0,0,0)
$$
where $v$ on the right hand side is the intersection point of the triple of
edges and it is understood that we only sum over such triples of edges 
which are incident at a common vertex. Moreover,
\be \label{13}
\hat{O}_{e,e',e^\dprime}(0,0,0)
=\frac{1}{8}\epsilon_{ijk}X^i_{e^\dprime} X^j_{e'} X^k_e \mbox{ and }
X^i_e:=X^i(h_e(0,1):=
\mbox{tr}(\tau_i h_e(0,1)\frac{\partial}{\partial h_e(0,1)}
\ee
is a right invariant vector field\footnote{This important observation,
which is fundamental for all that follows, is due to Abhay Ashtekar and 
Jurek Lewandowski} in 
the $\tau_i$ direction of $su(2)$,
that is, $X(hg)=X(h)$. The factor of $1/8$ in (\ref{13}) comes from the 
fact that at $t=t'=t^\dprime=0$ we get only one-sided functional derivatives
of the various edges involved ($m(0,0,0)=1/8$, see above). We also have 
extended the values of the sign function to include $0$ which takes care 
of the possibility that one has triples of edges with linearly dependent
tangents.\\
The final step consists in choosing $\Delta=\Delta'$ and taking the square 
root of the modulus. We replace the sum over all triples incident a common 
vertex $\sum_{e,e',e^\dprime}$ by a sum over all 
vertices followed by a sum over all triples incident at the same vertex
$\sum_{v\in V(\gamma)}\sum_{e\cap e'\cap e^\dprime=v}$.
Now, for small enough $\Delta$ and given $p$, at most one 
vertex contributes, that is, at most one of $\chi_\Delta(v,p)\not=0$ 
because all vertices have finite separation. Then we can 
take the relevant $\chi_\Delta(p,v)=\chi_\Delta(p,v)^2$ out of the square 
root and take the limit which results in 
\ba \label{14}
\hat{V}(R)_\gamma&=&\int_R d^3p \widehat{\sqrt{\det(q)(p)}}_\gamma 
=\int_R d^3p \hat{V}(p)_\gamma \nonumber\\
\hat{V}(p)_\gamma&=&\ell_p^3\sum_{v\in V(\gamma)} 
\delta^{(3)}(p,v)\hat{V}_{v,\gamma} \nonumber\\
\hat{V}_{v,\gamma}&=&\sqrt{|\frac{i}{3!\cdot 8} (\frac{3}{4})^3
\sum_{e,e',e^\dprime\in E(\gamma),e\cap e'\cap e^\dprime=v}
s(e,e',e^\dprime) q_{e e' e^\dprime}|} \nonumber\\
q_{e e' e^\dprime} &=& \epsilon_{ijk} X^i_e X^j_{e'} X^k_{e^\dprime}
\ea
where we could switch the order of the $X$'s because a triple contributes 
only if the corresponding edges are distinct and so the $X$'s commute.\\
Expression (\ref{14}) is the final expression for the volume operator and
coincides up to a factor $\sqrt{27/8}$ with the expression found in
\cite{5,9} while it is genuinely different from the one found in 
\cite{4} as pointed out in \cite{9}. Note that the final expression is 
manifestly diffeomorphism covariant. Although the procedure of adapting 
the limiting to a given triple of edges is somewhat non-standard there is an
argument in favour of such a procedure : the discussion in lemma (\ref{la1}) 
reveals that any other regularization which would result in a finite 
contribution for the case where $s(e,e' e^\dprime)$ is zero would necessarily
depend on the higher order intersection characteristics of a triple of edges.
However, since such a quantity is not diffeomorphism covariant which is 
inacceptable, the
dependence must be trivial, that is, a constant (this is precisely what 
happens in (\ref{4}) according to (\ref{9}). While there is no a priori 
{\em kinematical} reason to prefer one operator over the other, there is a 
{\em dynamical} reason :
the volume operator can be seen as an essential ingredient in the 
regularization of the Lorentzian Wheeler-DeWitt constraint operator
\cite{2,3}. The fact that (\ref{14}) contains $s(e,e',e^\dprime)$ is the
precise reason for this operator to be anomaly-free ! Therefore, the 
regularization leading to (\ref{14}) is the one singled out by the 
dynamics of the theory !

\section{Spectral Analysis}

In this section we summarize a number of results due to Abhay Ashtekar and 
Jurek Lewandowski obtained in the series of papers \cite{18,5,9}. 
We have included them here for the sake of self-containedness of the 
present paper. More specifically,
the cylindrical consistency of the family of volume operators 
$\hat{V}_\gamma$ as 
defined in \cite{18} and in the present paper, the self-adjointness of 
the corresponding projective limit and the fact 
that the spectrum of the volume operator is discrete was first observed in 
\cite{9}. 
The proof of these properties given in \cite{9} uses methods developed in 
\cite{18,1,5,9}. \\
\\
This section is subdivided into three parts. First we prove  
that the
family of operators derived in (\ref{14}) defines a linear unbounded 
operator on $\cal H$. Next we show that the operator is symmetric, 
positive semi-definite and admits self-adjoint extensions and finally we 
show that its 
spectrum is discrete and that the operator so defined is anomaly-free.

\subsection{Cylindrical Consistency}

What we have obtained in (\ref{14}) is a family of operators
$(\hat{V}(R)_\gamma,D_\gamma)_{\gamma\in\Gamma}$. That is not enough to 
show that this family of cylindrical projections ``comes from" a linear 
operator on $\cal H$. As proved in \cite{18}, for this to be the case
we need to check that whenever $\gamma\subset\gamma'$ then\\
1) $p_{\gamma\gamma'}^\ast D_\gamma\subset D_{\gamma'}$ where 
$p_{\gamma\gamma'}$ is the restriction from $\gamma'$ to $\gamma$.
This condition makes sure that the operator defined on bigger graphs can 
be applied to functions defined on smaller graphs.\\ 
2) $(\hat{V}(R)_{\gamma'})_{|\gamma}=\hat{V}(R)_\gamma$, this is the 
condition of cylindrical consistency and says that the operator on bigger
graphs equals the operator on smaller graphs when restricted to functions
thereon.\\
A graph $\gamma\subset\gamma'$ can be obtained from a bigger graph $\gamma'$
by a finite series of steps consisting of the following basic ones :\\
i) remove an edge from $\gamma'$,\\
ii) join two edges $e',e^\dprime$, such that $e'\cap e^\dprime$ is a point of
analyticity, to a new edge $e=e'\circ (e^\dprime)^{-1}$\\
iii) reverse the orientation of an edge.\\
Clearly, a dense domain for $\hat{V}(R)_\gamma$ is 
given by $D_\gamma:=\mbox{Cyl}_\gamma^3(\agb)$. This choice trivially
satisfies requirement 1) since functions which just do not depend on 
some arguments or only on special combinations $h_e=h_{e'} h_{e^\dprime},
h_{e'}=h_e^{-1}$ are still thrice continuously differentiable if the 
original function was (here we have used the fact that $SU(2)$ is a Lie 
group, that is, group multiplication and taking inverses is an analytic 
map).\\
Next, let us check cylindrical consistency.
Consider first the case i) that $\gamma$ does not depend on an edge $e$
on which $\gamma'$ does. Then clearly $X^i_e f_\gamma=0$ for any function
cylindrical with respect to $\gamma$ and so in the sum over triples over 
vertices in (\ref{14}) the terms involving $e$ drop out.\\ 
Next consider 
the case ii). If $e=e'\circ (e^\dprime)^{-1}$ is an edge of $\gamma$ and 
$e',e^\dprime$ are edges of $\gamma'$ where $v:=e'\cap e^\dprime$ is a point 
of analyticity for $\gamma$ while for $\gamma'$ it is not, then while $v$ 
is a 
vertex for $\gamma'$ it is only a pseudo-vertex for $\gamma$ and so in 
$\hat{V}(D)_\gamma$ there 
is no term corresponding to $v$. On the other hand, since the vertex $v$ is a 
pseudo vertex for $\gamma$ it is in particular only two-valent and so 
the corresponding term in $\hat{V}(D)_{\gamma'}$ drops out. Likewise, 
if $v$ is a vertex for $\gamma$ at which the outgoing edge $e$ is incident,
then from right invariance of the vector field we have $X_e=X_{e'\circ
(e^\dprime)^{-1}}=X_{e'}$ and so at vertices that belong to both $\gamma$ 
and $\gamma'$ the corresponding vertex operators coincide.\\
Finally, case iii) is actually excluded by our unambiguous choice of 
orientation. \\
We conclude that there exists an operator $(\hat{V}(R),D)$ on $\cal H$ 
which is densely defined on $D=\mbox{Cyl}^3(\agb)$.

\subsection{Symmetry, Positivity and Self-Adjointness}

Notice that the vector field $iX_e$ is symmetric on ${\cal H}_\gamma$,
the completion of $\mbox{Cyl}_\gamma^1(\agb)$ with respect to 
$\mu_{0,\gamma}$, $e$ an edge of $\gamma$, because the Haar measure is
right invariant. It follows from the explicit expression (\ref{14})
in terms of the $iX_e$ that all the projections $\hat{V}(R)_\gamma$ are 
symmetric. In this special case (namely, the volume operator leaves the
space $D_\gamma$ invariant) this is enough to show that $\hat{V}(R)$ is 
symmetric on $D$.\\ 
Furthermore, all $\hat{V}(R)_\gamma$ are positive semi-definite by 
inspection so that $\hat{V}(R),D$ is a densely defined, positive semidefinite
and symmetric operator. It follows that it has self-adjoint extensions,
for instance its Friedrich extension.

\subsection{Discreteness and Anomaly-freeness}

The operator $\hat{V}(D)$ has the important property that it leaves 
the dense subset $\mbox{Cyl}_\gamma^\infty)(\agb)\subset \cal H$
invariant, separately for each $\gamma\in\Gamma$. Moreover, recall that 
$\cal H$ possesses a convenient basis, called the spin-network basis
in the sequel \cite{21,22,23}.\\
Let us recall the most important features of that basis. Given a graph
$\gamma$, let us associate with each edge $e\in E(\gamma)$ an spin
quantum number $j_e>0$. Also, with each vertex $v\in V(\gamma)$ we associate
a certain contraction matrix $c_v$ as follows : if $\pi_j$ is the 
irreducible representation of $SU(2)$ corresponding to spin $j$ then 
form the tensor product $\otimes_{e\in E(\gamma)}\pi_{j_e}(h_e)$
and for each vertex $v$ contract the indices $A$ of the matrices 
$\pi_{j_e}(h_e)_{AB}$ for all $e$ incident at $v$ in a gauge invariant
fashion. We obtain a gauge invariant function 
$T_{\gamma,\vec{j},\vec{c}}(A)$, called a spin-network function. It turns out
that given $\gamma,\vec{j}$ there are only a finite number of linearly
independent $\vec{c}$ compatible with $\gamma,\vec{j}$. Moreover, one can
show that spin-network functions defined on different graphs are orthogonal
and even on the same graph they are orthogonal whenever the vectors $\vec{j}$
don't equal each other. It follows that there is an orthonormal basis
$$
<T_{\gamma,\vec{j},\vec{c}},T_{\gamma',\vec{j}',\vec{c}'}>
=\delta_{\gamma,\gamma'}\delta_{\vec{j},\vec{j}'}\delta_{\vec{c},\vec{c}'}\;.
$$
Now it is obvious that the operator $\hat{V}(R)$ leaves the finite 
dimensional vector space $U_{\gamma,\vec{j}}$ spanned by spin-network states
compatible with $\gamma,\vec{j}$ invariant. The matrix
\be \label{15}
(V(R)_{\gamma\vec{j}})_{\vec{c},\vec{c'}}:=
<T_{\gamma,\vec{j},\vec{c}}|\hat{V}(R)|T_{\gamma,\vec{j},\vec{c}'}>
\ee
is therefore finite-dimensional, positive semi-definite and symmetric. 
The task of computing its eigenvalues therefore becomes a problem in 
linear algebra ! \\
Next, since from (\ref{14})
\be \label{16}
\hat{V}(R)_\gamma=\ell_p^3\sum_{v\in V(\gamma)\cap R} \hat{V}_{v,\gamma}
\ee
and since $\hat{V}_{v,\gamma}$ involves only those $e\in E(\gamma)$ with
$v\in e$, we find that $\hat{V}_{v,\gamma}$ can only change the entry
$c_v$ in $\vec{c}$. In other words, 
$[\hat{V}_{v,\gamma},\hat{V}_{v',\gamma}]=0$ and each $\hat{V}_{v,\gamma}$
can be diagonalized separately.\\
Finally, since the spins $j_e$ only take discrete values it follows that
${\cal H}_\gamma$ has a countable basis and the spectrum that $\hat{V}(R)$
attains on $D_\gamma$ is therefore pure point. Let us check whether this 
is the complete spectrum. Assume it were not and let $\hat{P}$ the 
spectral projection on the rest of the spectrum (the existence of the 
spectral projections relies on the fact that $\hat{V}(R)$ is self-adjoint
and not only symmetric). It follows that $u=\hat{P}v$ is orthogonal to
$D_\gamma$ where $v$ is any vector in ${\cal H}_\gamma$. But $D_\gamma$
is dense in ${\cal H}_\gamma$ and so we find for every $\epsilon>0$ a
$\phi\in D_\gamma$ with $||u-\phi||<\epsilon$. Now we have from orthogonality
$\epsilon^2>||u-\phi||^2=||u||^2+||\phi||^2>||u||^2$ and so $u=0$. This 
shows that the complete spectrum is already attained on $D_\gamma$.
It is purely discrete as well\footnote{in the physical 
sense that it is attained on a countable basis so that the eigenvalues 
only comprise a countable set. In a mathematical sense one would need to 
check that there are no accumulation points and no eigenvalues of infinite
multiplicity. To settle this question it would be enough to show that
the formula for the eigenvalue is an unbounded function of
$j:=\sum_{e\in E(\gamma)} j_e$ because every countably infinite set
of $j$'s corresponding to different choices of $\vec{j}$ diverges. 
This is one possible future application of the explicit matrix element
formulae which we derive in the next section.}. \\
Last, we wish to show that the volume operators are anomaly free (given 
the fact that we have largely adapted our regularization to a graph, 
this statement is far from trivial). By this we mean the following : given 
any two open sets 
$R_1,R_2\subset\Sigma$ we have vanishing Poisson brackets $\{V(R_1),V(R_2)\}
=0$ because the functionals $V(R)$ depend on the momentum variable $E^a_i(x)$
only. Now, given a function $f$ cylindrical with respect to a graph $\gamma$,
it is not at all obvious any more that $[\hat{V}(R_1),\hat{V}(R_2)]f=0$
for any such $f$. Fortunately, given the above characterization of the 
spectrum, the 
commutator can be easily proved to vanish on cylindrical 
functions. To see this, note that the above results imply that if we choose
any region $R(\gamma)$ such that $\gamma\subset R(\gamma)$ then there exists
an eigenbasis of ${\cal H}_\gamma$ of $\hat{V}(R(\gamma))$. Now consider 
any region $R$. Since all regions are open by construction, all regions
fall into equivalence classes with respect to $\gamma$ : $R,R'$ are 
equivalent
if they contain the same vertices of $\gamma$ (any vertex either has a 
neighbourhood which lies completely inside $R$ or it lies outside). Therefore
any two $\hat{V}(R),\hat{V}(R')$ differ at most by some of the 
$\hat{V}_{v,\gamma}$ all of which are contained in the 
expression for $\hat{V}(R(\gamma))$.
Since the $\hat{V}_{v,\gamma}$ commute the eigenbasis of $\hat{V}(R(\gamma)$
is a simultaneous eigenbasis of all $\hat{V}_{v,\gamma}$ for all $v\in 
V(\gamma)$ and so this eigenbasis is a simultaneous eigenbasis of all
$\hat{V}(R)_{\gamma}$. Since all ${\cal H}_\gamma$ are orthogonal, we have
a simultaneous eigenbasis for all $\hat{V}(R)$.\\
While it is in general not enough to verify that two self-adjoint, unbounded 
operators 
commute on a dense domain (rather, by definition, we have to check that
the associated spectral projections commute) in our case we are done because
the spectral projections {\em are} the projections on the various 
$D_\gamma$ because the point spectrum is already the complete spectrum.
Thus we have verified that the commutator algebra mirrors the classical
Poisson algebra.

\section{The complete set of matrix elements}

In this final section we are going to compute (\ref{15}) in a spin-network
basis for each $\gamma,\vec{j},\vec{c},\vec{c}',R$. The advantage of
our approach as compared to the ones proposed already in the literature 
\cite{10} will be obvious : in \cite{10} one works with an overcomplete and
non-orthogonal set of states. This prevents one from giving a unified 
treatment as it is done here and only allowed one to find a formula for the
eigenvalues in a few simple cases. More generally one has to check all 
the time the linear independence of the states and can compute the matrix
elements only in a case by case analysis.\\
In the next subsection we show in more detail a result which follows 
immediately from \cite{1,5,9} : namely, that the problem of computing the 
matrix elements of the volume operator reduces to that of computing the 
matrix elements of a homogenous polynomial of degree three of spin operators
for a spin system. 

This opens access to powerful techniques well-known
from the quantum theory of angular momentum for spin systems. Then we will
show how the actual computation can be done by viewing the problem as 
a task in spin recoupling theory and that all the matrix elements can be 
written as algebraic functions of $6j$ symbols for which a closed 
formula exists (Racah formula). Finally we will diagonalize the volume 
operator for an infinite number of (very special) graphs.

\subsection{The left regular representation and the spin-network 
representation}

Consider a spin-network state $T_{\gamma,\vec{j},\vec{c}}(A)$. This is a 
state for the connection representation $L_2(\agb,d\mu_0)$ in the sense
that the (equivalence classes modulo gauge transformations of) connection 
operators are diagonal in this representation. Notice that a spin-network
state is gauge invariant so that it can be viewed alternatively as a 
function of $[A]$, the gauge equivalence class of $A$. Here we will
use the fact that $d\mu_0$ extends to $\ab$ \cite{20} and reduces to
the measure indicated in the introduction when integrating gauge invariant
functions.\\
Consider the usual Dirac (generalized) states $\delta_A$ defined by
$\delta_A(A'):=\delta_{\mu_0}(A,A')$ where $\delta_{\mu_0}$ is the 
$\delta$ distribution with respect to $\mu_0$ \cite{23} and 
$A,A'\in\ab$. We define the {\em spin-network representation} by
\be \label{17}
\delta_A(T_{\gamma,\vec{j},\vec{c}}):=\int_\ab d\mu_0(A')
\overline{\delta_A(A')} T_{\gamma,\vec{j},\vec{c}}(A')
=T_{\gamma,\vec{j},\vec{c}}(A)=:<A|\gamma,\vec{j},\vec{c}> \;.
\ee 
So far this is only notation. The advantage of this notion will become 
transparent upon proving that the states $|\gamma,\vec{j},\vec{c}>$
are nothing else than the usual angular momentum states associated 
with an abstract spin system. \\
To see this, let $v$ be a vertex of $\gamma$ and $e_1,..,e_n$ the 
edges of $\gamma$ incident at $v$ and we have chosen orientations 
such that all of them are outgoing at $v$. The spin-network state
$T_{\gamma,\vec{j},\vec{c}}(A)$ can be written
\ba \label{18}
T_{\gamma,\vec{j},\vec{c}}(A)
&=&(c_v)_{m_1..m_n}\prod_{i=1}^n (\pi_{j_{e_i}}(h_{e_i}(A)))_{m_i m'_i}
(M_v)_{m'_1..m'_n} \nonumber\\
&=& \mbox{tr}(c_v\cdot\otimes_{i=1}^n \pi_{j_{e_i}}(h_{e_i}(A))\cdot M_v)
\ea
where $c_v$ is the vertex contractor corresponding to $v$ which contracts 
the indices corresponding to the starting points of the $e_i$ and $M_v$ 
contracts the indices corresponding to the endpoints (recall that 
the holonomy is a path ordered exponential with the smallest parameter
value to the left). Note that $c_v$ is $\Co-$valued while $M_v$ still depends
on the rest of the graph.\\
Consider a local gauge transformation at $v$, that is, $h_{e_i}\to g 
h_{e_i}$ for some $g\in SU(2)$. Then (\ref{18}) becomes
$\mbox{tr}(c_v\cdot\otimes_{i=1}^n \pi_{j_{e_i}}(g)\cdot
\otimes_{i=1}^n\pi_{j_{e_i}}(h_{e_i}(A))\cdot M_v)$. In order for this to 
be gauge invariant, notice that the tensor product representation 
$\otimes_i \pi_{j_{e_i}}(g)$ can be written as a direct sum of 
orthogonal irreducible representations $\pi_j(g)$, and so we just need to 
choose $c_v$ to be 
proportional to one of the various equivalent (but orthogonal) trivial
representations $\pi_0(g)=\pi_0(1)$ contained in that decomposition
(see \cite{23} for more details) in order for (\ref{18}) to be gauge
invariant. This is precisely what a spin-network state is.\\
The point is now that 
\be \label{19}
\prod_{i=1}^n (\pi_{j_{e_i}}(h_{e_i}(A)))_{m_i m'_i}
=:<A|(m_1,j_1),..,(m_n,j_n);m'_1,..,m'_n>
\ee
transforms {\em precisely} as the usual angular momentum eigenstates
$$
|(m_1,j_1),..,(m_n,j_n);k_1,..,k_m>:=\prod_{i=1}^n |m_i,j_i;k_1,..,k_m>$$ 
of an abstract spin system
of $n$ spins $j_1,..,j_n$ under rotations (see, e.g., \cite{24}). 
This is because $|(m,j);k>$ transforms as
$\hat{U}(g)|(m,j);k>=\pi_j(g)_{m,m'}|(m',j);k>$ by definition, where
$\hat{U}(g)$ is a unitary representation of the rotation group.
Here $k_1,..,k_m$ are some additional quantum numbers of observables 
commuting with the operators $\hat{J_i}$ which in our case 
coincide with the $m_1',..,m_n'$. In other words, the left regular
representation $L_g f(h_e)=f(g h_e)$ that we are dealing with in the
connection representation can be seen to be equivalent with the
representation $\hat{U}(g)$ of the spin-network representation.  \\
Let us now couple the spins $j_1,..,j_n$ to resulting spin $j=0$ and 
denote a particular state with zero angular momentum by $|0>$. We claim that
up to a constant
\be \label{20}
(c_v)_{m_1.. m_n}=<(m_1,j_1),..,(m_n,j_n);k_1,..,k_m|0>
\ee
where the inner product is now to be understood in the angular momentum 
Hilbert space of the abstract spin system. In other words, the $c_v$ are 
just usual Clebsh-Gordan coefficients. The proof is 
easy, we have\\ 
$(c_v)_{m_1..m_n} \prod_{i=1}^n \pi_{j_i}(g)_{m_i m'_i}(g)$\\
$=<\hat{U}(g)((m_1',j_1),..,(m_n',j_n);k_1,..,k_m)|0>$\\
$=<(m'_1,j_1),..,(m'_n,j_n);k_1,..,k_m|\hat{U}(g^{-1}) 0>$\\
$=(c_v)_{m_1'..m_n'}$\\
where we have used unitarity of the group and of the representation and that
$\hat{U}(g)|0>=|0>$ is rotation invariant. The factor of proportionality
is seen to equal unity because $c_v^2=1$ and 
since $\sum_{m_1.. m_n} |(m_1,j_1)..(m_n,j_n);k_1..k_m>
<(m_1,j_1)..(m_n,j_n);k_1..k_m|=1$ due to completeness and Wigner-Eckart
theorem \cite{24}.\\
Next, consider the self-adjoint right invariant vector fields $Y_e:=-i/2 
X_e$.
We have $[X^i,X^j]=-2\epsilon^{ijk} X^k$ and so $[Y^i,Y^j]=i\epsilon^{ijk}
Y^k$ which is the usual angular momentum algebra. It follows that (we use
a dual interpretation of the $SU(2)$ indices $A,B,C,..$ which we may 
choose to take values $\pm 1/2$ and so can be interpreted to be the 
eigenvalues of the 3-component of angular momentum)
\ba 
\label{21} && |j=1/2,C>
Y_e^j h_e(A)_{CD}= <j=1/2,C| (-i/2)(\tau_j h_e(A))_{CD} \nonumber\\
&=&(i/2)<(\tau_j)_{EC}(1/2,C)|  h_e(A))_{ED} \nonumber\\
&=&(i/2)\frac{d}{dt}_{t=0}<\hat{U}(\exp(t\tau_j)(1/2,E)|  
h_e(A))_{ED}=<1/2,E|J_j h_e(A)_{ED}
\ea
and so by multilinearity of the tensor product representation we find
due to self-adjointness of $J_I$ (set $Y_I:=Y_{e_I}$)
\ba \label{22}
&& Y^j_I T_{\gamma,\vec{j},\vec{c}}(A) 
\nonumber\\
&=&\frac{i}{2}\frac{d}{dt}_{t=0} 
<\vec{j},\vec{m}><A|\gamma,\vec{j},\vec{m};\vec{m}'
|\hat{U}(-\exp(t\tau_j))_I 0>
(M_v)_{m_1',..m_n'}
\nonumber\\
&=&
<\vec{j}_v,\vec{m}><A|\gamma,\vec{j}_v,\vec{m};\vec{m}'|J^j_I 0>
(M_v)_{m_1',..m_n'} \ea
where $\vec{j}_v$ comprises the components of $\vec{j}$ which correspond to
the edges incident at $v$.\\
Since spin-network states are thus characterized by the various 
zero-angular momentum eigenstates $|0>$ of the abstract spin system, there 
is a one-to-one correspondence between the possible choices of
$c_v$ and $|0>$. So let $|0>$ correspond to $c_v$ and $|0'>$ to 
$c_v'$.
Let $p(\vec{Y})$ be any polynomial of the $Y_I^i$ corresponding to 
the vertex $v$ then we have for its matrix elements 
\ba \label{23}
&& <T_{\gamma,\vec{j},\vec{c}}|p(\vec{Y})|T_{\gamma,\vec{j},\vec{c}'}>
\nonumber\\
&=& \int_\agb d\mu_0(A)
\overline{<\vec{j}_v,\vec{m}|0><A|\gamma,\vec{j}_v,\vec{m};\vec{m}'> 
(M_v)_{m_1',..m_n'}}
<\vec{j}_v,\vec{\bar{m}}|p(\vec{J}) 0'>
\times\nonumber\\
&& \times
<A|\gamma,\vec{j}_v,\vec{\bar{m}};\vec{\bar{m}}'> 
(M_v)_{\bar{m}_1',..\bar{m}_n'}
\nonumber\\
&=& \overline{<\vec{j}_v,\vec{m}|0>}
<\vec{j}_v,\vec{m}|p(\vec{J})^\dagger 0'>
\prod_{v'\not=v}\delta_{c_{v'},c_{v'}'}\nonumber\\
&=& <0|p(\vec{J})^\dagger 
0'>\prod_{v'\not=v}\delta_{c_{v'},c_{v'}'}
=<p(\vec{J}) 0|0'>\prod_{v'\not=v}\delta_{c_{v'},c_{v'}'}
\;. 
\ea
Here we have used (\ref{22}) in the first step, the orthonormality of the
spin-network states in the second step and in the last step we have again
recalled that $\sum_{\vec{m}}|\vec{j}_v,\vec{m}><\vec{j}_v,\vec{m}|=1$
is the identity on the abstract angular momentum Hilbert space labelled
by the spins $\vec{j}_v$ (completeness, notice that we sum over $\vec{m}$ 
in (\ref{23})).\\ 
Expression (\ref{23}) is a huge simplification : in order to compute 
the matrix elements of the volume operator it is enough to compute
the matrix elements of the angular momentum operator polynomial
$\epsilon_{ijk} J^i_e J^j_{e'} J^k_{e^\dprime}$
between states that have zero angular momentum coming from
$\vec{j}_v$. This task is devoted to the next subsection.

\subsection{Recoupling Theory for $n$ angular momenta}

The first task is to label all the linearly independent states $|0>$
which correspond to the coupling of $n$ angular momenta $j_1,..,j_n$
to zero angular momentum. This is a well known problem in the 
quantum theory of angular momentum and runs under the name recoupling theory.
We review here what we need for our computation.\\
The states $|\vec{j},\vec{m}>$ are a complete basis for a spin system 
consisting of $n$ angular momenta. In this basis the $2n$ mutually 
commuting operators $(J_I^i)^2,J_I^3$ are diagonal with eigenvalues 
$j_I(j_I+1),m_I$. \\
We are obviously interested in a basis in which the square of the 
operator of total angular momentum $J:=J_1+..+J_n$ and its $3-$ component
are diagonal with eigenvalues $j(j+1),m$. In order to label that basis we 
need $2(n-1)$ more quantum numbers of mutually commuting operators and 
commuting with $J^2,J_3$. The $n$ operators $J_I^2$ are readily verified to
satisfy these conditions but we need $n-2$ more. In order to motivate our 
choice notice that $\hat{V}_{v,\gamma}$ is the square root of the modulus 
of $i(3/2)^3 \frac{1}{4} \sum_{I<J<K}\epsilon(e_I,e_J,e_K) q_{IJK}$ 
where 
$$
\frac{1}{4}q_{IJK}:=\frac{1}{8}\epsilon_{ijk}X^i_I X^j_J X^k_K=
[Y_I^i Y^i_J,Y^j_J Y^j_K]=\frac{1}{4}[(Y_{IJ}^i)^2,(Y_{JK}^j)^2]
$$
where $Y_{IJ}:=Y_I+Y_J$. Here we have used total antisymmetry of 
$\epsilon(e_I,e_J,e_K),\hat{q}_{IJK}$ and the fact that $[X_I,X_J]=0,
I\not= J$ in order to get rid of $3!$ and in order to sum only over $I<J<K$.
This observation\footnote{This fact that the third 
order polynomial $\mbox{tr}(X_I[X_J,X_K])$ in the $X$ can be written as a 
commutator of two second order polynomials $\mbox{tr}(X_I X_J)$ 
which are gauge invariant was
made independently also by Abhay Ashtekar and Jurek Lewandowski}   
urges us to study the operator \be \label{24}
\hat{q}_{IJK}:=[J_{IJ}^2,J_{JK}^2] \mbox{ where } 
J_{IJ}=J_I+J_J
\ee
and thus it would be convenient if we could work in a basis in which
the $J_{IJ}$ also were diagonal. Clearly we cannot achieve this 
for all $n(n-1)/2$ operators $J_{IJ}$ but we can work in different
bases adapted to $\hat{q}_{IJK}$ for a particular triple $I<J<K$ and 
expand them in terms of each other.\\ 
We will develop a theory of 
recoupling schemes here which is somewhat different from the one 
encountered in the standard literature and we also consider arbitrary $n$.
The following paragraphs will therefore be somewhat detailed.
\begin{Definition}
A coupling scheme based on a pair $1\le I<J\le n$ is an orthonormal basis 
$|\vec{g}(IJ),\vec{j},j,m>$, diagonalizing
besides $J^2,\;J_3,\;J_I^2,\;I=1,..,n$ the squares of the additional $n-2$ 
operators $G_2,..,G_{n-1}$ where :\\ 
$G_1:=J_I,\; G_2:=G_1+J_J,\;G_3:=G_2+J_1,\;G_4:=G_3+J_2,\;..,\;
G_{I+1}:=G_I+J_{I-1},\; G_{I+2}:=G_{I+1}+J_{I+1},\; G_{I+3}:=G_{I+2}+
J_{I+2},\;..,\;G_{J}:=G_{J-1}+J_{J-1},\; G_{J+1}:=G_J+J_{J+1},\;
G_{J+2}:=G_{J+1}+J_{J+2},\;..,\;G_n:=G_{n-1}+J_n=J$.\\
The vector $\vec{g}(IJ):=(g_2(IJ),..,g_{n-1}(IJ))$ denotes the 
eigenvalues of the squares of $G_2,..,G_{n-1}$.\\
We will call the coupling scheme based on the pair $(12)$ the standard basis.
\end{Definition}
Let us check that the $G_I$ satisfy the angular momentum algebra and that 
the 
squares of the operators $J,J_I,G_I$ and the operator $J^3$ are mutually 
commuting :
for this it will be enough to check that the operators for the standard 
basis are mutually commuting since there is a relabelling of indices such 
that any coupling scheme becomes the standard basis. \\
Since $[J_I,J_J]=0,I\not=J$ we find $[G_I^i,G_I^j]
=\sum_{J=1}^I [J_J^i,J_J^j]=i\epsilon_{ijk} \sum_{J=1}^I J_J
=i\epsilon_{ijk}G_I^k$ which implies that $G_I^2$ has spectrum
$g_I(g_I+1)$ with $g_I$ integral or half integral.\\
Let $I<J$, then
we can write $G_J=G_I+(J_{I+1}+..+J_J)=:G_I+G_I'$ so that $[G_I,G_I']=0$
because $G_I,G_I'$ involve disjoint sets of $J_K$'s and so
$[G_J^2,G_I^2]=[G_I^2+(G_I')^2,G_I^2]+
2(G_I')^i\{G_I^j[G_I^i,G_I^j]+[G_I^i,G_I^j]G_I^j\}
=2i(G_I')^i\epsilon_{ijk}\{G_I^j G_I^k+G_I^k G_I^j\}=0$.
The same argument can be used for the other commutators.\\
The virtue of this definition is the following : we have now a neat 
labelling of the vectors $|0>$ referred to in the previous subsection :
we can choose them to be the vectors of the standard basis corresponding 
to $j=0$, that is, they are just the vectors $|\vec{g}(12),\vec{j},j=0,m=0>$
($m=0$ is forced by $j=0$) and the different vertex contractors just 
correspond to the various 
possible values of the intermediate (or so-called re-coupling) vectors
$G_2,..,G_{n-1}$.
We will therefore compute the matrix elements of $\hat{q}_{IJK}$ in this
basis. In the sequel we will drop the indices $\vec{j},j=0,m=0$ since 
they are the same in every coupling scheme and since $\hat{q}_{IJK}$ 
commutes with $J^2,J_I^2,(J)^3$ matrix elements between vectors with 
different values of these operators therefore vanish anyway. We will
also write $\vec{g}$ instead of $\vec{g}(12)$. \\
After these long preparations we are now in the position to complete the 
derivation. We use completeness of the bases $|\vec{g}(IJ),\vec{j},j,m>$ and 
have 
\ba \label{25}
&&<\vec{g}|\hat{q}_{IJK}|\vec{g}'>\nonumber\\
&=& \sum_{\vec{g}(IJ)}g_2(IJ)(g_2(IJ)+1)
[<\vec{g}|\vec{g}(IJ)><\vec(g_{IJ})|J_{JK}^2|\vec{g}'>
-<\vec{g}|J_{JK}^2|\vec{g}_{IJ}><\vec{g}(IJ)|\vec{g}'>]
\nonumber\\
&=& 
\sum_{\vec{g}(IJ),\vec{g}(JK),\vec{g}^\dprime}
g_2(IJ)(g_2(IJ)+1)g_2(JK)(g_2(JK)+1)
<\vec{g}(IJ)|\vec{g}^\dprime><\vec{g}(JK)|\vec{g}^\dprime>\times\nonumber\\
&&\times
[<\vec{g}(IJ)|\vec{g}><\vec{g}(JK)|\vec{g}'>
-<\vec{g}(JK)|\vec{g}><\vec{g}(IJ)|\vec{g}'>]
\ea
where we have used that the so-called {\em $3nj-$symbols}  
$<\vec{g}(IJ)|\vec{g}(JK)>$
are real (also $<\vec{g}(IJ)|\vec{g}(KL)>$ for any $I<J,K<L$ are real).
To see this, insert a unit in terms of the basis 
$|\vec{j},\vec{m}>$ which allows us to write the $3nj-$symbol in terms
of Clebsh-Gordan coefficients which are real \cite{24}. It is understood 
that in (\ref{25}) and in what follows the summation extends only over 
the allowed values of $\vec{g}$ etc. which are purely fixed by the values 
of $j_1,..,j_n$.\\
The task left is to write the $3nj-$symbols of the form
$<\vec{g}(IJ)|\vec{g}'>$ in terms of known 
quantities.
We will derive below a formula which expresses it in terms of the $6j$
symbol for which a closed formula (the Racah formula \cite{24}) exists.\\
We will denote
$\vec{g}(IJ)=(g_2(j_I,j_J),g_3(g_2,j_1),..,g_{I+1}(g_I,j_{I-1}),
g_{I+2}(g_{I+1},j_{I+1}),..,$\\$ g_J(g_{J-1},j_{J-1}),g_{J+1}(g_J,j_{J+1}),
..g_{n-1}(g_{n-2},j_{n-1)})$ where the notation $j^\dprime(j,j')$ means 
that we couple $J,J'$ to resulting spin $j^\dprime$ of $J^\dprime$ thereby
keeping explicit track of the coupling scheme. The following two
elementary lemmata will be crucial in computing the $3nj-$symbols.
\begin{Lemma} \label{la2}
$<\vec{g}(IJ)|\vec{g}'>$\\
$=<g_2(j_I,j_J),g_3(g_2,j_1),..,g_{I+1}(g_I,j_{I-1}),
g_{I+2}(g_{I+1},j_{I+1}),..,g_J(g_{J-1},j_{J-1})|$\\
$|g'_2(j_1,j_2),g'_3(g'_2,j_3),..,g'_J(g'_{J-1},j_J)>
\delta_{g_J,g'_J}..\delta_{g_{n-1},g'_{n-1}}$ where on the right hand side 
we consider a spin system consisting of spins $J_1,..,J_J$ and total spin
quantum number $g_J$.
\end{Lemma}
Proof :\\
Since $G_{J}(12)^2,..,G_{n-1}(12)^2$ are diagonal on both vectors it is
immediately clear that the $3nj-$symbol must be proportional to the Kronecker
delta $\delta_{g_K,g'_K}$ for $K=J,..,n-1$. To get the factor of 
proportionality we expand both vectors in terms of the basis 
$|\vec{j},\vec{m}>$ and Clebsh-Gordan coefficients. To do this, note that
by definition\\
$<g_2(j_1,j_2),..,g_{n-2}(g_{n-3},j_{n-2});g_{n-1} m-m_n,j_n m_n$\\
$|g_2(j_1,j_2),..,g_{n-2}(g_{n-3},j_{n-2}),g_{n-1}(g_{n-2},j_{n-1}),
j(g_{n-1},j_n),m>$\\
$=<g_{n-1}(g_{n-2},j_{n-1})m-m_n,j_n m_n|j,m;g_{n-1},j_n>$
where on the right hand side we have the standard Clebsh-Gordan coefficient.
Here\\ 
$|g_2(j_1,j_2),..,g_{n-2}(g_{n-3},j_{n-2});g_{n-1} m-m_n,j_n m_n>$\\
$:=|g_2(j_1,j_2),..,g_{n-2}(g_{n-3},j_{n-2}),g_{n-1}(g_{n-2},j_{n-1}), 
m-m_n>\otimes |j_n m_n>$. We can now iterate this procedure and obtain\\
$|g_2'(j_1,j_2),..,g_{n-2}'(g_{n-3}',j_{n-2}),
g_{n-1}'(g_{n-2}',j_{n-1}),j(g_{n-1}',j_n),m>$\\
$=\sum_{m_J+..+m_n=m}
<g_{n-1}' m-m_n, j_n m_n|j m> <g_{n-2}' m-m_n-m_{n-1},j_{n-1} 
m_{n-1}|g_{n-1}' m-m_n>..$\\
$.. <g_J' m_J, j_{J+1} m_{J+1}|g_{J+1}', m_J+m_{J+1}>$\\
$|g_2'(j_1,j_2),..,g_J'(g_{J-1}',j_J),m_J>\otimes|j_{J+1},m_{J+1}>
\otimes..\otimes |j_n,m_n>$. Similarily we have\\
$|g_2(j_I,j_J),..,g_{n-2}(g_{n-3},j_{n-2}),g_{n-1}(g_{n-2},
j_{n-1}),j(g_{n-1},j_n),m>$\\
$=\sum_{m_J+..+m_n=m}<g_{n-1} m-m_n, j_n m_n|j m>$ \\
$<g_{n-2} m-m_n-m_{n-1},j_{n-1} m_{n-1}|g_{n-1} 
m-m_n>..<g_J m_J, j_{J+1} m_{J+1}|g_{J+1}, m_J+m_{J+1}>$\\
$|g_2(j_I,j_J),..,g_J(g_{J-1},j_{J-1}),m_J>\otimes|j_{J+1},m_{J+1}>
\otimes..\otimes|j_n,m_n>$. Thus we have the inner product\\
$<\vec{g}(IJ)|\vec{g}'>=\sum_{m_J+..+m_n=m}$\\
$<g_2(j_I,j_J),..,g_J(g_{J-1},j_{J-1}),m_J|
g_2'(j_1,j_2),..,g_J'(g_{J-1}',j_J),m_J>\times$\\
$\times <g_{n-1}' m-m_n, j_n m_n|j m> <g_{n-2}' m-m_n-m_{n-1},j_{n-1} 
m_{n-1}|g_{n-1}' m-m_n>..<g_J' m_J, j_{J+1} m_{J+1}|g_{J+1}', m_J+m_{J+1}>
\times$\\ $\times
<g_{n-1} m-m_n, j_n m_n|j m> <g_{n-2} m-m_n-m_{n-1},j_{n-1} m_{n-1}|g_{n-1} 
m-m_n>..<g_J m_J, j_{J+1} m_{J+1}|g_{J+1}, m_J+m_{J+1}>$\\
$=<g_2(j_I,j_J),..,g_J(g_{J-1},j_{J-1}),\tilde{m}|
g_2'(j_1,j_2),..,g_J'(g_{J-1}',j_J),\tilde{m}>\times$\\
$\times\sum_{m_{J+1},..,m_n}
<g_{n-1}' m-m_n, j_n m_n|j m> <g_{n-2}' m-m_n-m_{n-1},j_{n-1} 
m_{n-1}|g_{n-1}' m-m_n>..<g_J' m_J, j_{J+1} m_{J+1}|g_{J+1}', m_J+m_{J+1}>
\times$\\$\times
<g_{n-1} m-m_n, j_n m_n|j m> <g_{n-2} m-m_n-m_{n-1},j_{n-1} m_{n-1}|g_{n-1} 
m-m_n>..<g_J m_J, j_{J+1} m_{J+1}|g_{J+1}, m_J+m_{J+1}>$\\
$=<g_2(j_I,j_J),..,g_J(g_{J-1},j_{J-1}),\tilde{m}|
g_2'(j_1,j_2),..,g_J'(g_{J-1}',j_J),\tilde{m}>
\delta_{g_J,g'_J}..\delta_{g_{n-1},g'_{n-1}}$
where in the second step we have used the fact that the $3nj-$symbols 
are independent of $m$ (Wigner-Eckart theorem\footnote{Proof :
Let us expand $|\vec{g}(IJ)>=\sum_{\vec{g}(KL)}
<\vec{g}(KL);j,m|\vec{g}(IJ);j,m> |\vec{g}(KL);j,m>$ where 
we assume that $m<j$ w.l.g. Now apply the ladder operator $J^\dagger$
to both sides of this equation. We find $J^\dagger|\vec{g}(MN);j,m>=
c(j,m)|\vec{g}(MN);j,m+1>$ for any $K<L$ where the coefficient only depends
on $j,m$. Comparing coefficients we find
$<\vec{g}(KL);j,m|\vec{g}(IJ);j,m>=<\vec{g}(KL);j,m+1|\vec{g}(IJ);j,m+1>$,
that is, the $3nj-$symbols are rotation invariant. $\Box$}) so that
$\tilde{m}$ is arbitrary and in the last step we have 
used the sum rule for the Clebsh-Gordan coefficients, that 
is, $\sum_{m_1+m_2=m}(<j_1 m_1,j_2 m_2|j m>)^2=1$.\\ 
$\Box$
\begin{Lemma} \label{la3}
$<g_2(j_1,j_2),..,g_K(g_{K-1},j_K),g_{K+1}(g_K,j_{K+1},g_{K+2}(g_{K+1},j_{K+2})
;m|$\\
$|g_2(j_1,j_2),..,g_K(g_{K-1},j_K),g_{K+1}'(g_K,j_{K+2}),
g_{K+2}'(g_{K+1}',j_{K+1});m>$\\
$=<g_{K+1}(g_K,j_{K+1}),g_{K+2}(g_{K+1},j_{K+2});m|
g_{K+1}'(g_K,j_{K+2}),g_{K+2}(g_{K+1}',j_3);m>$.
\end{Lemma}
Proof :\\
Again we expand \\
$|g_2,..,g_{K+2};m>=\sum_{m_{K+1},m_{K+2}}$\\
$<g_K m-m_{K+1}-m_{K+2},j_{K+1} m_{K+1}|g_{K+1} m-m_{K+2}>
<g_{K+2} m-m_{K+2},j_{K+2} m_{K+2}|g_{K+2} m-m_{K+2}>\times$\\
$\times |g_2,..,g_k;m-m_{K+1}-m_{K+2}>\otimes |j_{K+1} m_{K+1},j_{K+2}
m_{K+2}>$ and \\
$|g_2,..,g_K,g_{K+1}',g_{K+2};m>=\sum_{m_{K+1},m_{K+2}}$\\
$<g_K m-m_{K+1}-m_{K+2},j_{K+2} m_{K+2}|g_{K+1}' m-m_{K+1}>
<g_{K+2} m-m_{K+1},j_{K+1} m_{K+1}|g_{K+2} m-m_{K+1}>\times$\\
$\times |g_2,..,g_k;m-m_{K+1}-m_{K+2}>\otimes |j_{K+1} m_{K+1},j_{K+2}
m_{K+2}>$. We find for the inner product of these states due to the 
normalization just\\
$\sum_{m_{K+1},m_{K+2}}
<g_K m-m_{K+1}-m_{K+2},j_{K+1} m_{K+1}|g_{K+1} 
m-m_{K+2}>\times$\\ $\times <g_{K+2} m-m_{K+2},j_{K+2} m_{K+2}|g_{K+2} 
m-m_{K+2}>\times$\\ $\times<g_K m-m_{K+1}-m_{K+2},j_{K+2} m_{K+2}|g_{K+1}' 
m-m_{K+1}> <g_{K+2} m-m_{K+1},j_{K+1} m_{K+1}|g_{K+2} m-m_{K+1}>$\\
$\equiv 
<g_{K+1}(g_K,j_{K+1}),g_{K+2}(g_{K+1},j_{K+2});m|
g_{K+1}'(g_K,j_{K+2}),g_{K+2}(g_{K+1}',j_3);m>$ as claimed.\\
$\Box$\\
We are now ready to reduce out the general $3nj-$symbol. We assume
that $g_K=g_K',K=J,..,n-1$ as otherwise we get zero by lemma
\ref{la2}. Upon repeatedly using the completeness relations of the 
coupling scheme bases we first shift $j_J$ and then $j_I$ to the right
in the matrix element
\ba \label{26}
&& <\vec{g}(IJ),\vec{g}'>\nonumber\\
&=_{\ref{la2}} &
<g_2(j_I,j_J),g_3(g_2,j_1),..,g_{I+1}(g_I,j_{I-1}),
g_{I+2}(g_{I+1},j_{I+1}),..,g_J(g_{J-1},j_{J-1})|
\nonumber\\
&&| 
g_2'(j_1,j_2),g_3'(g_2',j_3),..,g_{J-1}'(g_{J-2}',j_{J-1}),g_J(g_{J-1}',j_J)>
\nonumber\\
&=_{\ref{la2}}& \sum_{h_2}
<g_2(j_I,j_J),g_3(g_2,j_1),g_4,..,g_J|
h_2(j_I,j_1),g_3(h_2,j_J),g_4,..,g_J>\nonumber\\
&& <h_2(j_I,j_1),g_3(h_2,j_J),g_4,..,g_J|
g_2'(j_1,j_2),g_3'(g_2',j_3),..,g_{J-1}',g_J>
\nonumber\\
&=_{\ref{la2}}& \sum_{h_2}
<g_2(j_I,j_J),g_3(g_2,j_1)|h_2(j_I,j_1),g_3(h_2,j_J)>\times\nonumber\\
&& \times \sum_{h_3}
<h_2(j_I,j_1),g_3(h_2,j_J),g_4(g_3,j_2),g_5,..,g_J|
h_2(j_I,j_1),h_3(h_2,j_2),g_4(h_3,j_J),g_5,..,g_J>\nonumber\\
&& <h_2(j_I,j_1),h_3(h_2,j_2),g_4(h_3,j_J),g_5,..,g_J|
g_2'(j_1,j_2),g_3'(g_2',j_3),..,g_{J-1}',g_J>
\nonumber\\
&=_{\ref{la2},\ref{la3}}& 
\sum_{h_2} 
<g_2(j_I,j_J),g_3(g_2,j_1)|h_2(j_I,j_1),g_3(h_2,j_J)>\times\nonumber\\
&&\times \sum_{h_3} <g_3(h_2,j_J),g_4(g_3,j_2)|h_3(h_2,j_2),g_4(h_3,j_J)>
\times\nonumber\\
&&\times <h_2(j_I,j_1),h_3(h_2,j_2),g_4(h_3,j_J),g_5,..,g_J|
g_2'(j_1,j_2),g_3'(g_2',j_3),..,g_{J-1}',g_J>
\nonumber\\
&=& ...\nonumber\\
&=&
\sum_{h_2} 
<g_2(j_I,j_J),g_3(g_2,j_1)|h_2(j_I,j_1),g_3(h_2,j_J)>\times\nonumber\\
&& \times
\sum_{h_3} <g_3(h_2,j_J),g_4(g_3,j_2)|h_3(h_2,j_2),g_4(h_3,j_J)>..
\nonumber\\
&&\sum_{h_I} 
<g_I(h_{I-1},j_J),g_{I+1}(g_I,j_{I-1})|h_I(h_{I-1},j_{I-1}),g_{I+1}(h_I,j_J)>
\times \nonumber\\ 
&& \times \sum_{h_{I+1}} 
<g_{I+1}(h_I,j_J),g_{I+2}(g_{I+1},j_{I+1})|h_{I+1}(h_I,j_{I+1}),
g_{I+2}(h_{I+1},j_J)>\times\nonumber\\
&&\times\sum_{h_{I+2}} 
<g_{I+2}(h_{I+1},j_J),g_{I+3}(g_{I+2},j_{I+2})|h_{I+2}(h_{I+1},j_{I+2}),
g_{I+3}(h_{I+2},j_J)>..
\nonumber\\
&&\sum_{h_{J-1}}
<g_{J-1}(h_{J-2},j_J),g_J(g_{J-1},j_{J-1})|h_{J-1}(h_{J-2},j_{J-1}),
g_J(h_{J-1},j_J)>\times
\nonumber\\
&&\times
<h_2(j_I,j_1)h_3(h_2,j_2),..,h_I(h_{I-1},j_{I-1}),h_{I+1}(h_I,j_{I+1}),..,
h_{J-1}(h_{J-2},j_{J-1}),g_J(h_{J-1},j_J)|\nonumber\\
&&| g_2'(j_1,j_2),..,g_I'(g_{I-1}',j_I),
g_{I+1}'(g_I',j_{I+1}),..,g_J'(g_{J-1}',j_J)>\;.
\ea
Now the last matrix element in the last line of (\ref{26}) is different 
from zero only for $h_K=g_K',K=I,..,J-1$ by lemma \ref{la2} (we have 
$g_J=g_J'$ already) and in 
that case reduces by the same reasoning to
\ba \label{27}
&&
<h_2(j_I,j_1)h_3(h_2,j_2),..,h_{I-1}(h_{I-2},j_{I-2}),g_I'(h_{I-1},j_{I-1})|
g_2'(j_1,j_2),..,g_I'(g_{I-1}',j_I)>
\nonumber\\
&=&\sum_{k_2} 
<h_2(j_I,j_1),h_3(h_2,j_2)|k_2(j_1,j_2),h_3(k_2,j_I)>\times\nonumber\\
&&\times \sum_{k_3} <h_3(k_2,j_I),h_4(h_3,j_3)|k_3(k_2,j_3),h_3(k_3,j_I)>..
\nonumber\\
&& \sum_{k_{I-1}} 
<h_{I-1}(k_{I-2},j_I),h_I(h_{I-1},j_{I-1})|k_{I-1}(k_{I-2},j_{I-1}),
g_I'(k_{I-1},j_I)>\times
\nonumber\\
&& \times
<k_2(j_1,j_2),k_3(k_2,j_3),..,k_{I-1}(k_{I-2},j_{I-1}),g_I'(k_{I-1},j_I)|
g_2'(j_1,j_2),g_3'(g_2',j_3),..,g_I'(g_{I-1}',j_I)>\nonumber\\
&&
\ea 
and the last matrix element in the last line of (\ref{27}) is just given by 
$\prod_{K=2}^{I-1} \delta_{k_K,g_K'}$ so that the summation drops out.
So, altogether, there are only the $I-3$ summation variables 
$h_2,..,h_{I-1}$.\\
Equations (\ref{26}), (\ref{27}) demonstrate that we manage to reduce
the $3nj-$symbol under investigation to a polynomial of matrix elements 
of the structure
\ba \label{28}
&& \mbox{Type I : }
<j_{12}(j_1,j_2),j(j_{12},j_3)|j_{13}(j_1,j_3),j(j_{13},j_2)> 
\mbox{ or } 
\nonumber\\
&& \mbox{Type II : }
<j_{12}(j_1,j_2),j(j_{12},j_3)|j_{23}(j_2,j_3),j(j_{23},j_1)> \;.
\ea
In fact, all matrix elements in (\ref{26}) are of type I and all matrix 
elements in (\ref{27}) are of type I except for the first factor 
after the first equality which is of type II. \\
The point is now that both matrix elements are related to the $6j-$symbol
\cite{24}. The $6j-$symbol is defined through the type II matrix element,
namely
\ba \label{29}
&&<j_{12}(j_1,j_2),j(j_{12},j_3)|j_{23}(j_2,j_3),j(j_{23},j_1)> 
\nonumber\\
&=:&\sqrt{(2j_{12}+1)(2j_{23}+1)}(-1)^{j_1+j_2+j_3+j}
\left\{ \begin{array}{ccc}
j_1 & j_2 & j_{12}\\
j_3 & j   & j_{23}
\end{array} \right\} \;.
\ea
In order to relate the type I matrix element to the $6j-$symbol we need to
use the following identity for the CG-coefficients
\be \label{30}
<j_1 m_1,j_2 m_2|jm>=(-1)^{j_1+j_2-j}<j_2 m_2,j_1 m_1|j m>
\ee
which is a consequence of the choice of phases for the CG coefficients in 
order to have them real valued \cite{24}. Let us now set
$j_1':=j_2,j_2':=j_1,j_3':=j_3,j_{12}':=j_{12},j_{23}':=j_{13},j':=j$
then we get
$$
<j_{12}(j_1,j_2),j(j_{12},j_3)|j_{13}(j_1,j_3),j(j_{13},j_2)>=
<j_{12}'(j_2',j_1'),j'(j_{12}',j_3')|j_{23}'(j_2',j_3'),j'(j_{23}',j_1')> 
$$
which is almost the standard form except for the wrong order of $j_1',j_2'$
in the first entry. Using the expansion\\
$|j_{12}'(j_2',j_1'),j'(j_{12}',j_3');m'>$\\
$\sum_{m_1+m_2+m_3=m'} <j_2' m_2,j_1' m_1|j_{12}' m_1+m_2>
<j_{12}' m_1+m_2, j_3' m_3|j' m'> |j_1' m_1,..,j_3' m_3'>$
and (\ref{30}) we therefore find\\
$
<j_{12}'(j_2',j_1'),j'(j_{12}',j_3')|j_{23}'(j_2',j_3'),j'(j_{23}',j_1')>$\\
$=(-1)^{j_1'+j_2'-j_{12}'}
<j_{12}'(j_1',j_2'),j'(j_{12}',j_3')|j_{23}'(j_2',j_3'),j'(j_{23}',j_1')> 
$ 
and so we have for the type I matrix element\\
\ba \label{31}
&& <j_{12}(j_1,j_2),j(j_{12},j_3)|j_{13}(j_1,j_3),j(j_{13},j_2)>
\nonumber\\
&=& (-1)^{j_1+j_2-j_{12}}
\sqrt{(2j_{12}+1)(2j_{13}+1)}(-1)^{j_1+j_2+j_3+j}
\left\{ \begin{array}{ccc}
j_2 & j_1 & j_{12}\\
j_3 & j   & j_{13}
\end{array} \right\} \;.
\ea
Finally, for the benefit of the reader we note the Racah formula \cite{24}
\ba \label{32}
&&\left\{ \begin{array}{ccc}
j_1 & j_2 & j_{12}\\
j_3 & j   & j_{23}
\end{array} \right\}
=\Delta(j_1,j_2,j_{12})\Delta(j_1,j,j_{23})\Delta(j_3,j_2,j_{23})
\Delta(j_3,j,,j_{12})w
\left\{ \begin{array}{ccc}
j_1 & j_2 & j_{12}\\
j_3 & j   & j_{23}
\end{array} \right\}, \nonumber\\
&&
\Delta(a,b,c)=\sqrt{\frac{(a+b-c)!(a-b+c)!(-a+b+c)!}{(a+b+c+1)!}}
\nonumber\\
&& 
w
\left\{ \begin{array}{ccc}
j_1 & j_2 & j_{12}\\
j_3 & j   & j_{23}
\end{array} \right\}
=\sum_n (-1)^n (n+1)! \times \nonumber\\
&&\times [(n-j_1-j_2-j_{12})!(n-j_1-j-j_{23})!(n-j_3-j_2-j_{23})!
(n-j_3-j-j_{12})!]^{-1}\times
\nonumber\\
&&\times [(j_1+j_2+j_3+j-n)!(j_2+j_{12}+j+j_{23}-n)!
(j_{12}+j_1+j_{23}+j_3-n)!]^{-1} 
\ea
and the sum extends over all positive integers such that no factorial in 
the denominator has a negative argument. In conclusion, formulae (\ref{25}), 
(\ref{26}), (\ref{27}) provide us
with the general expression for the matrix elements of the operator
$\hat{q}_{IJK}$ in the standard basis where it is understood that
all occurring matrix elements are known in terms of $j_1,..,j_n$ via
(\ref{29}), (\ref{31}), (\ref{32}). \\
Let us call $M_{v,IJK}(\vec{g},\vec{g}'):=<\vec{g}|\hat{q}_{IJK}|
\vec{g}'>$ then we have the matrix 
\be \label{33}
M_v(\vec{g},\vec{g'})=(\frac{3}{2})^3 \frac{i}{4}\sum_{I<J<K}\epsilon(I,J,K)
M_{v,IJK}(\vec{g},\vec{g}')
\ee
which by inspection of (\ref{25}) is a Hermitean matrix of a special kind,
namely, it is of the form of $i$ times a real-valued skew matrix. It's 
eigenvalues 
therefore come in pairs $\pm\lambda,\lambda>0$ or are zero. The eigenvalues
of the positive semidefinite matrix $M_v^\dagger M_v$ are therefore
$|\lambda|^2$ or $0$ where $\lambda$ is an eigenvalue of $M_v$.
We conclude that the eigenvalues of $\hat{V}_{v,\gamma}$ are given by
$\ell_p^3 \sqrt{|\lambda|}$ or $0$ where both eigenvectors of $M_v$
corresponding to eigenvalues $\pm\lambda$ produce the same eigenvalue
$\ell_p^3 \sqrt{|\lambda|}$ of $\hat{V}_{v,\gamma}$. We see that the problem
of diagonalizing the volume operator becomes equivalent to diagonalizing
the matrices $M_v$ which can be done for each $v\in V(\gamma)$ separately.\\
In the next subsection we will derive the spectrum in the most simple 
cases.\\
Remark :\\
We have nowhere made use of the fact that we are only interested in
gauge invariant functions. Therefore, our results carry over, word by word,
to the case of non-gauge invariant functions provided we use the notion of an
extended spin-network state.
\begin{Definition}
An extended spin-network state is a state of the form 
$T_{\gamma,\vec{j},\vec{c}}$ as before just that the vertex contractors 
now are Clebsh-Gordan coefficients of the form\\
$<j_1 m_1,..,j_n m_n|j_1,..,j_n,g_2,..,g_{n-1},j>$ where $j$ maybe different
from zero.
\end{Definition} 
This is important for applications in quantum gravity \cite{2,3} and for 
the length operator \cite{6} where 
we need the action of the volume operator on non-gauge invariant functions 
in an intermediate step.

\subsection{Examples}

Given a vertex $v$ and a vector of spins $\vec{j}$ colouring the edges 
incident at $v$, let us denote by $d(\vec{j},v)$ the number of linearly 
independent vectors $|\vec{g}>$ compatible with $\vec{j}$. This number
is the dimension of a vector space of vertex contractors associated with
$v$ and it is given purely algebraically as the number of linearly
independent trivial representations which appear in the decomposition into
irreducibles of the tensor product representation 
$\pi_{j_1}\otimes..\otimes\pi_{j_n}$. One can determine this number for
each case at hand by repeated application of the usual Clebsh-Gordan theorem
$\pi_j\otimes\pi_{j'}=\pi_{j+j'}\oplus\pi_{j+j'-1}\oplus..\oplus
\pi_{|j-j'|}$ and counting all the trivial representation of spin $0$ 
that appear. Notice that the sum is direct so that all the representations
are orthogonal and therefore linearly independent. What one would like to 
have is a closed formula for $d(\vec{j},v)$ but that seems to be a 
complicated combinatorical problem. To our knowledge, such a dimension 
formula is only known for the case of an $n-fold$ tensor product of
fundamental representations $j_1=..=j_n=1/2$ for the groups 
$GL(N),U(N),SU(N)$ \cite{25}. We will not deal with this problem in the 
present paper.\\ 
First of all, whenever $d(\vec{j},v)=1$
then $0$ is the only possible eigenvalue (this observation
was made first in \cite{10} for the case of a trivalent graph in which the
vertex contractors span only a one-dimensional vector space). 
More generally, whenever $d(\vec{j},v)$ is odd, we know that $0$ is an
eigenvalue of multiplicity at least one.\\
Since $M_v$ is a $d\times d$ skew matrix its characteristic polynomial
has the structure $t^k(t^2-\lambda_1^2)..(t^2-\lambda_{(d-k)/2}^2)=t^k 
p_{(d-k)/2}(t^2)$ where $0\le k\le d$ for $d$ even and $1\le k\le d$ for 
$d$ odd. It follows from Galois theory that we can determine the 
spectrum by quadratures exactly in general whenever $d(\vec{j},v)\le 9$
and in fortunate cases (whenever $0$ has multiplicity at least $d-8$)
up to arbitrarily high dimension. In the most general case, however,
a numerical evaluation is the only possible approach.\\
We consider the first non-trivial cases $d(\vec{j},v)\le 9$ for an arbitrary 
valence $n$ of the vertex.
We may label the spins in such a way that $0<j_1\le..\le j_n$. We may 
parametrize the situation by the non-negative number $k:=j_1+..+j_{n-1}-j_n$
(there is no contractor for $k<0$).
If $k$ is low valued then it is clear from Clebsh-Gordan theory that
$d(\vec{j},v)$ will be low valued. Specifically, if $k=0$ then the first 
$n-1$ 
spins have to add up to maximum spin and thus $d=1$, the spectrum is only
the point $0$ irrespective of the value of $n$. In the following we will 
assume that $2j_1\ge k$, otherwise the discussion requires a case 
division.\\ 
If $k=1$ then one sees readily that $d=n-1$ : the only 
irreducible representations contained in $j_1\otimes j_2$ that can possibly 
add up with $j_3,..,j_{n-1}$ to $j_n$ are $j_1+j_2,j_1+j_2-1$. Repeating 
the argument, the assertion follows.\\
Let in general $c_k(n)$ be the number of representations with weight
$j_1+..+j_n-k$ 
contained in $j_1\otimes..\otimes j_n$ for $j_{n+1}=j_1+..+j_n-k$ to
couple to angular momentum $0$ where
$j_1\le,..,\le j_n$ and where we assume that $j_2-j_1\le j_1+j_2-k$,
that is $k\le 2j_1$. We wish to show that $c_k(n)=(n+k-2,k)$ where 
$(n,k)$ is the usual binomial coefficient. Since $c_0(n)=1$ as we 
showed above, the induction is started. Now 
suppose we know all coefficients up to $k$ for each $n$ and start deriving
those for $k+1$. We can assume that we know them already up to $n-1$ since
$c_{k}(2)=1$ for all $k$ is trivial to see since we have made the 
assumption $2j_1\ge k$. Then we have the recursion $c_{k+1}(n-1)+c_k(n)
=c_{k+1}(n)$ which follows from the fact that if we have the tensor product
$j_1\otimes..\otimes j_{n-1}$ then there are $c_l(n-1)$ representations 
with weight $j_1+..+j_{n-1}-l,0\le l\le k+1$. But if we now form the tensor
product with all of them with $j_n$, then the number of times that 
$j_1+..+j_n-k$ arises from those with weight $j_1+..+j_{n-1}-l,0\le l\le k$, 
which is given by $c_k(n)$, is equal to 
the number of times that $j_1+..+j_n-(k+1)$ arises from them. In addition 
we have $c_{k+1}(n-1)$ terms coming from the representation 
$j_1+..+j_{n-1}-(k+1)$ itself. It is now readily verified that 
$c_k(n)$ solves the recursion.\\
Notice that $n+1$ is the valence of the vertex. So we get back the result 
that in the tri-valent case we have $d=c_k(2)=(k,k)=1$
We can now ask for which combinations of $k,n$ we have $c_k(n)\le 9$.
Notice that $c_k(3)=k+1$ :\\
Case $k=0$ : $n$ arbitrary.\\
Case $k=1$ : $n-1\le 9$ so $n\le 10$.\\
Case $k=2$ : $n(n-1)/2\le 9$ so $n\le 4$.\\
Case $3\le k\le 8$ : $n\le 3$.\\
Case $k>8$ : $n\le 2$.\\
Reverse question (only $n\ge 2$ makes sense since we need valence at least
$3$ for the volume operator not to vanish trivially) :\\
Case $n=2$ : $k$ arbitrary.\\
Case $n=3$ : $k+1\le 9$ so $k\le 8$.\\
Case $n=4$ : $(k+1)(k+2)/2\le 9$ so $k\le 2$.\\
Case $5\le n\le 10$ : $k\le 1$.\\
Case $n>10$ : $k=0$.\\
So, for instance we can treat the general 4-valent case under the restriction
that $j_4=j_1+j_2+j_3-k,0\le k\le 8$ for $j_3\ge j_2\ge j_1\ge k/2$ in 
which case $d=k+1$. As the formulae already get quite involved, we will 
only discuss the cases $k=1,2,3$ explicitly.\\
Notice first of all that due to gauge-invariance 
$J_1+J_2+J_3+J_4|0>=0$ so that we can substitute for $J_4$. Thus for instance
$\hat{q}_{124}|0>=-(\hat{q}_{121}+\hat{q}_{122}+\hat{q}_{123})|0>$ and it is
easily verified that the operator identity $\hat{q}_{122}+\hat{q}_{121}=0$
holds. It follows that 
$$
M_v=(\frac{3}{4})^3\frac{i}{4}[\epsilon(1,2,3)-\epsilon(1,2,4)+
\epsilon(1,3,4)-\epsilon(2,3,4)]\hat{q}_{123}=:\sigma\hat{q}_{123}
$$
which simplifies the discussion tremendously since we just need to 
compute the matrix elements of the single operator $\hat{q}_{123}$.
We have to compute the $k(k+1)/2$ matrix elements
$$
<j_{12}|\hat{q}_{123}|j_{12}'>=[j_{12}(j_{12}+1)-j_{12}'(j_{12}'+1)]
\sum_{j_{23}}j_{23}(j_{23}+1)
<j_{12}|j_{23}><j_{12}'|j_{23}>
$$
for $j_{12},j_{12}'=j_1+j_2-l,\;l=0,..,k$ and the sum extends over
$j_{23}=j_2+j_3-l,\;l=0,..,k$ which requires the computation of 
$(k+1)^2$ $6j$ symbols for $j_{123}=j=j_1+j_2+j_3-k=j_4$.
The computation is rather tedious. We will display only a few intermediate
calculational results before giving the eigenvalue $\lambda_k$.\\
Let $j_{12}:=j_1+j_2,j_{23}:=j_2+j_3,J=2(j_1+j_2+j_3)$.\\
Case $k=1$.\\
The $6j$ symbols are 
\ba \label{33a}
<j_{12}|j_{23}>&=&-<j_{12}-1|j_{23}-1> 
=\sqrt{\frac{j_1 j_3}{j_{12}j_{23}}} \nonumber\\
<j_{12}-1|j_{23}>&=& <j_{12}|j_{23}-1> 
=\sqrt{\frac{j_2 J}{2j_{12}j_{23}}} 
\ea 
and so we have the only non-vanishing matrix element
\be \label{34}
<j_{12}|\hat{q}_{123}|j_{12}-1>=4\sqrt{j_1 j_2 j_3(j_1+j_2+j_3)} \;.
\ee
The corresponding eigenvalue of the volume operator is
\be \label{35}
\lambda_1=2\ell_p^3\sqrt{|\sigma|}\root 4\of{j_1 j_2 j_3(j_1+j_2+j_3)}\;.
\ee
Case $k=2$ :\\
The $6j$ symbols are 
\ba \label{36}
<j_{12}|j_{23}>&=&\sqrt{\frac{j_1(2j_1-1)j_3(2j_3-1)}{j_{12}(2j_{12}-1)
j_{23}(2j_{23}-1)}} \nonumber\\
<j_{12}-1|j_{23}>&=&\sqrt{2\frac{(2j_1-1)j_2 j_3(J-1)}{j_{12}(2j_{12}-2)
j_{23}(2j_{23}-1)}} \nonumber\\
<j_{12}-2|j_{23}>&=&\sqrt{\frac{j_2(2j_2-1)(J-1)(J-2)}{(2j_{12}-1)(2j_{12}-2)
j_{23}(2j_{23}-1)}} \nonumber\\
<j_{12}|j_{23}-1>&=&\sqrt{2\frac{j_1 j_2 (2j_3-1)(J-1)}{j_{12}(2j_{12}-1)
j_{23}(2j_{23}-2)}} \nonumber\\
<j_{12}|j_{23}-2>&=&\sqrt{\frac{j_2(2j_2-1)(J-1)(J-2)}{j_{12}(2j_{12}-1)
(2j_{23}-1)(2j_{23}-2)}} \nonumber\\
<j_{12}-1|j_{23}-1>&=&\frac{(2j_2-1)(J-1)-(2j_1-1)(2j_3-1)}
{2\sqrt{j_{12}(2j_{12}-2)j_{23}(2j_{23}-2)}} \nonumber\\
<j_{12}-2|j_{23}-1>&=&-\sqrt{2\frac{(2j_1-1)(2j_2-1) 
j_3(J-2)}{(2j_{12}-1)(2j_{12}-2) j_{23}(2j_{23}-2)}} \nonumber\\
<j_{12}-1|j_{23}-2>&=&-\sqrt{2\frac{j_1(2j_2-1) 
(2j_3-1)(J-2)}{j_{12}(2j_{12}-2)(2j_{23}-1)(2j_{23}-2)}} \nonumber\\
<j_{12}-2|j_{23}-2>&=&2\sqrt{\frac{j_1(2j_1-1) 
j_3(2j_3-1)}{(2j_{12}-1)(2j_{12}-2)(2j_{23}-1)(2j_{23}-2)}}
\ea
and we find then the matrix elements
\ba \label{37}
<j_{12}|\hat{q}_{123}|j_{12}-1>&=&
\sqrt{\frac{2j_1 2j_2 
(2j_3-1)(J-1)}{2(2j_{12}-1)(2j_{12}-2)}}\frac{1}{2(2j_{23}-1)}\times\nonumber\\
& & \times [(2j_1-1)[6j_3+2 j_{23}-1]+(2 j_2-1)[3 J +2 j_{23}-7]]\nonumber\\
<j_{12}|\hat{q}_{123}|j_{12}-2>&=&0\nonumber\\
<j_{12}-1|\hat{q}_{123}|j_{12}-2>&=&
\sqrt{\frac{(2j_1-1)(2j_2-1) 
2j_3(J-2)}{2(2j_{12})(2j_{12}-1)}}\frac{1}{2(2j_{23}-1)}\times\nonumber\\
& & \times [(J-1)[6j_2+2 j_{23}-1]+(2 j_3-1)[6 j_1 -2 j_{23}+1]]
\nonumber\\
&&
\ea
and since the eigenvalues of any $3\times 3$ skew matrix with entries
$a,b,c$ are given by $0,\pm i\sqrt{a^2+b^2+c^2}$
we find the eigenvalues $0,\lambda_2$ where
\be \label{38}
\lambda_2=\ell_p^3\sqrt{|\sigma|}\root 4\of{<j_{12}|\hat{q}_{123}|j_{12}-1>^2
+<j_{12}-1|\hat{q}_{123}|j_{12}-2>^2}\;.
\ee
Case $k=3$ :\\
We will refrain from displaying the $6j-$symbols explicitly in terms of 
the $j_I$, rather we consider them as given through the formulae of the 
previous section. Now it is easy to check that the eigenvalues of a 
$4\times 4$ skew matrix with entries $a,b,c,d,e,f$ are given by 
$i$ times \\
$t=\pm \frac{1}{\sqrt{2}}\sqrt{a^2+b^2+c^2+d^2+e^2+f^2\pm
\sqrt{[(a+f)^2+(c+d)^2+(e-f)^2]\times}}$\\
$\overline{\overline{\times[(a-f)^2+(c-d)^2+(e+f)^2]}}$\\
and notice that the argument of the inner square root can be written as 
$(a^2+b^2+c^2+d^2+e^2+f^2-4(af+cd-eb)^2$ which shows that $t$ is 
purely real. Substituting the matrix elements 
$<j_{12}-l|\hat{q}_{123}|j_{12}-l'>,\;0\le l<l,\le 3$ for $a,b,c,d,e,f$
we obtain
\be \label{39}
\lambda_3=\ell_p^3\sqrt{|\sigma|}\sqrt{|t|}\;.
\ee
The reader should now have a feeling of how to compute the spectrum 
analytically. The formulae derived can serve for analytical estimates
as well as the starting point for the implementation of an appropriate 
computer code for an algebraic manipulation programme.\\
\\
\\
\\
{\large Acknowledgements}\\
\\
The author is indebted to Abhay Ashtekar and Jurek Lewandowski for 
communicating their results on the volume operator prior to publication.\\
This research project was 
supported in part by DOE-Grant DE-FG02-94ER25228 to Harvard University.


\begin{thebibliography}{99}

\parskip -5pt

\bibitem{1} A. Ashtekar, J. Lewandowski, D. Marolf, J. Mour\~ao, T.
Thiemann, ``Quantization for diffeomorphism invariant theories 
of connections with local degrees of freedom", Journ. Math. Phys.
{\bf 36} (1995) 519-551

\bibitem{2} T. Thiemann, ``Anomaly-free formulation of non-perturbative,
four-dimensional Lorentzian quantum gravity", 
Harvard Preprint HUTMP-96/B-350, Physics Letters B (in press)

\bibitem{3} T. Thiemann, ``Quantum Spin Dynamics (QSD)", 
Harvard Preprint HUTMP-96/B-351\\
T. Thiemann, ``Quantum Spin Dynamics (QSD) II", 
Harvard Preprint HUTMP-96/B-352

\bibitem{4} C.\ Rovelli, L.\ Smolin, ``Discreteness of volume and 
area in quantum gravity'' Nucl. Phys. B {\bf 442} (1995) 593, Erratum :
Nucl. Phys. B {\bf 456} (1995) 734

\bibitem{5} A. Ashtekar, J. Lewandowski, ``Quantum Theory of Geometry I:
Area Operators", Preprint CGPG-96/2-4, gr-qc/9602046  \\
``Quantum Theory of Geometry II: Volume Operators", in preparation

\bibitem{6} T. Thiemann, ``The length operator in canonical quantum gravity",
            Harvard Preprint HUTMP-96/B-354

\bibitem{7} T. Thiemann, ``The ADM Hamiltonian Operator for canonical 
quantum gravity", Harvard University Preprint

\bibitem{8} T. Thiemann, ``A regularization of canonical Yang-Mills quantum
theory", Harvard University Preprint

\bibitem{9} J. Lewandowski, ``Volume and Quantization", Potsdam Preprint,
gr-qc/9602035

\bibitem{10} R. Loll, ``Spectrum of the volume operator in quantum gravity",
Preprint DFF 235/11/95, gr-qc/9511030

\bibitem{11} A. Ashtekar, in ``Mathematics and general relativity", AMS,
Providence, 1987

\bibitem{12} . Barbero, Phys. Rev. D {\bf 51} (1995) 5507

\bibitem{13} R. Giles, Phys. Rev. {\bf D8} (1981) 2160

\bibitem{14} R. Gambini, A. Trias, Nucl. Phys. {\bf B278}, 436 (1986).

\bibitem{15} A. Ashtekar and C.J. Isham,
Class. \& Quan. Grav. {\bf 9}, 1433 (1992).

\bibitem{16} A. Ashtekar and J. Lewandowski, ``Representation
theory of analytic holonomy $C^\star$ algebras'', in {\it Knots and
quantum gravity}, J. Baez (ed), (Oxford University Press, Oxford 1994).

\bibitem{17} J. Baez, Lett. Math. Phys. {\bf 31}, 213 (1994);
``Diffeomorphism invariant generalized measures on the space of
connections modulo gauge transformations", hep-th/9305045,
in the Proceedings of the conference on quantum topology, D. Yetter
(ed) (World Scientific, Singapore, 1994).

\bibitem{18} A. Ashtekar and J. Lewandowski, ``Differential
geometry on the space of connections via graphs and projective
limits'', Journ. Geo. Physics {\bf 17} (1995) 191

\bibitem{19} D. Marolf and J. M. Mour\~ao, ``On the support of the
Ashtekar-Lewandowski measure'',  Commun. Math. Phys. {\bf 170} (1995)
583-606

\bibitem{20} A. Ashtekar and J. Lewandowski, J. Math. Phys. {\bf 36}, 2170
(1995). 

\bibitem{21} C.\ Rovelli, L.\ Smolin,
``Spin networks and quantum gravity'' pre-print CGPG-95/4-1. 

\bibitem{22} J.\ Baez, ``Spin network states in gauge theory",
Adv. Math. (in press); ``Spin networks in non-perturbative quantum
gravity,'' pre-print gr-qc/9504036.

\bibitem{23} T.\ Thiemann, ``The inverse Loop Transform", preprint
HUTMP-95/B-346, hep-th/9601105     

\bibitem{24} A. R. Edmonds, ``Angular momentum in quantum mechanics",
Princeton University Press, Princeton, New Jersey, 1974

\bibitem{25} H. Boerner, ``Darstellungen von Gruppen", Springer-Verlag,
Berlin, 1967

\end{thebibliography}
\end{document}